\documentclass[12pt, english]{iopart}
\usepackage{lmodern}

\usepackage[T1]{fontenc}
\usepackage[utf8]{inputenc}
\usepackage{rotating}
\expandafter\let\csname equation*\endcsname\relax
\expandafter\let\csname endequation*\endcsname\relax
\usepackage{amsmath}
\usepackage{amssymb}
\usepackage{graphicx}
\usepackage{esint}

\makeatletter
\providecommand{\tabularnewline}{\\}
\makeatother

\usepackage{babel}

\usepackage{xcolor}
\usepackage{comment}

\newcommand{\modification}[1]{#1}
\newcommand{\scheme}[3]{${#1}\overset{#3}{\longrightarrow}{#2}$}

\begin{document}
\global\long\def\kex#1{\left|\mathrm{X}_{#1}\right\rangle }
 \global\long\def\kph#1{\left|\mathrm{P}_{#1}\right\rangle }
 \global\long\def\kp#1{\left|\mathrm{p}_{#1}\right\rangle }

\global\long\def\bex#1{\left\langle \mathrm{X}_{#1}\right|}
 \global\long\def\bph#1{\left\langle \mathrm{P}_{#1}\right|}
 \global\long\def\bp#1{\left\langle \mathrm{p}_{#1}\right|}

\global\long\def\bld#1{\mathbf{#1}}
 \global\long\def\vk{\textbf{k}}

\title{Entanglement generation in microcavity polariton devices}

\author{L. Einkemmer}
\address{Department of Experimental Physics, University of Innsbruck, 6020
Innsbruck, Austria}

\author{Z. Vörös}
\ead{zvoros@uibk.ac.at}
\address{Department of Experimental Physics, University of Innsbruck, 6020
Innsbruck, Austria}

\author{G. Weihs}
\address{Department of Experimental Physics, University of Innsbruck, 6020
Innsbruck, Austria}

\author{S. Portolan}
\address{Institute of Atomic and Subatomic Physics, TU Wien, 1020 Wien, Austria}

\begin{abstract}
Entanglement generation in microcavity exciton-polaritons is an interesting
application of the peculiar properties of these half-light/half-matter
quasiparticles. In this paper we theoretically investigate their luminescence
dynamics and entanglement formation in single, double, and triple
cavities. We derive general expressions and selection rules for polariton-polariton
scattering. We evaluate a number of possible parametric scattering
schemes in terms of entanglement, and identify the ones that are experimentally
most promising. 
\end{abstract}
\maketitle

\section{Introduction}

Microcavity polaritons have come to the fore of semiconductor optics
research after Weisbuch et al. \cite{Weisbuch1992} demonstrated strong
coupling of the cavity photon and the quantum well exciton. In their
work, the photon in a single cavity was coupled with excitons of a
quantum well located at the anti-node of the electric field. Soon
afterwards, several groups demonstrated strong coupling in double
\cite{Armitage1998,Panzarini1999,Panzarini1999a,Panzarini1997,Stanley1994},
and triple cavities \cite{Diederichs2006,Diederichs2007a,Diederichs2007b}.
Beyond the fact that these structures would allow more exotic scattering
scenarios, the advantage of coupled cavities is that polariton-polariton
scattering can be studied on branches that are protected from the
exciton reservoir, or in other context, the excitation-induced dephasing,
and thus, one of the non-radiative polariton decay channels is removed.
In a different context, this was already pointed out in the work of
Ciuti \cite{Ciuti2004}, and in Pagel et al. \cite{Pagel2012a}.

The polariton dispersion relations (including polarization splitting) of coupled cavities were discussed in several
papers, including Armitage at al. \cite{Armitage1998}, Panzarini et al.
\cite{Panzarini1999,Panzarini1999a,Panzarini1997}, and Stanley et al. \cite{Stanley1994}, while the triple cavity case
was presented in Diederichs et al. \cite{Diederichs2006,Diederichs2007a,Diederichs2007b}. However, all of these papers
were concerned with the dispersion relations only, and, to some extent, their role in polariton-polariton scattering,
but the question of scattering with other quasi-particles was not addressed. Therefore, from these works alone, it would
not be clear whether a particular polariton-polariton scattering process is experimentally feasible: even if the process
is allowed, the resulting polaritons could be buried in a strong background of thermal polaritons. \modification{To
analyze this aspect we will extend the quantum Langevin approach introduced in \cite{Portolan2008} to double and triple
cavity configurations. This approach allows us to provide quantitative predictions of noise induced by thermally
scattered pump polaritons, which is expected to be the dominant source in experimentally realistic polariton devices. In
addition, in the context outlined above, we are able to compare the pump-induced photoluminescence of double cavity
schemes with the corresponding results for single cavity devices (which were considered in the literature before, see,
for example,  \cite{Portolan2008c}).}

In this paper, we present a detailed study of polariton scattering in multiple cavities, and identify the dominant
scattering processes \modification{with the aim of finding schemes that support the generation of
polarization-entangled light}. The paper is organized as follows. First, starting with a simple model,
we inspect the symmetry properties of polaritons in coupled and triple cavities. Based on these symmetries, we then
establish some general selection rules. Following this, we derive the equations governing polariton-polariton
scattering. An analytical solution and its discussion for the case of continuous-wave pumping is given in Section
\ref{sec:analytical-modeling}, and we measure entanglement of the produced light. The case of pulsed excitation is
detailed in Section \ref{sec:numerical-simulations}. Here we also compare the results to that obtained for the
analytical case, and show that meaningful conclusions can be drawn from the steady state solution. We close the paper
with some general remarks and a short outlook in Section \ref{sec:conclusion}. The technical details of our derivations
are outlined in the Appendix.

\section{The role of the symmetry\label{sec:symmetry}}

In the rest of the paper, we will apply the following notation: exciton,
and photon states located in a particular cavity are denoted by $\kex i$,
and $\kph i$, respectively, while we will use $\left|\mathrm{p}_{j\vk}\right\rangle $for
a polariton on the $j$th branch with momentum $\hbar\mathbf{k}$.
(The exact definition of the polariton branches will is stated below.)

\begin{figure}
\centering
\includegraphics[width=0.65\textwidth]{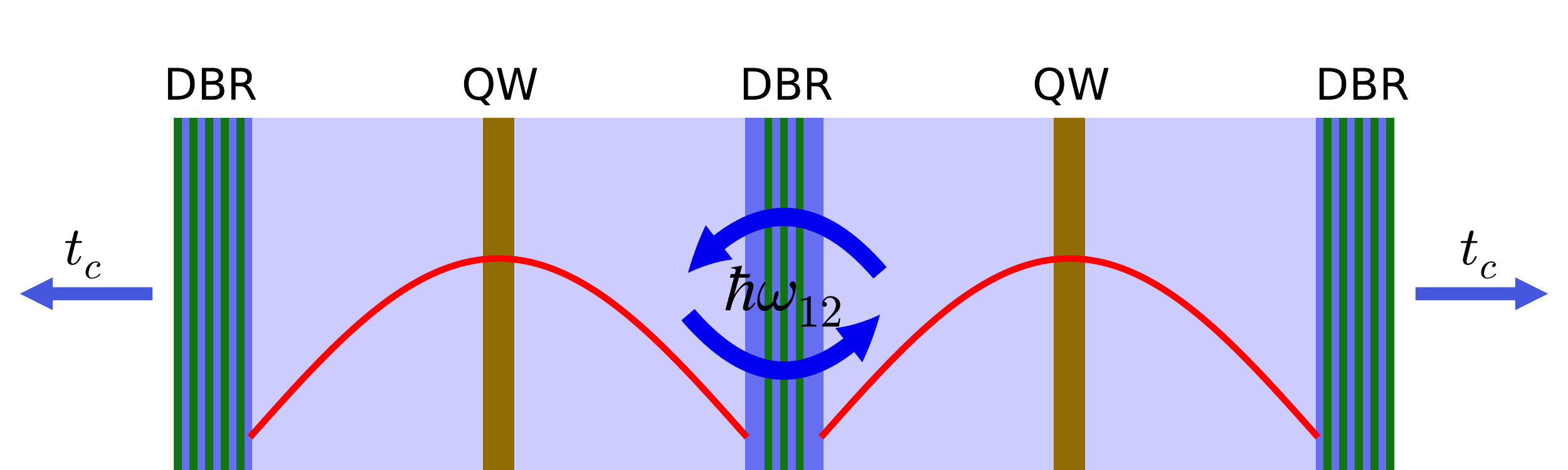}
\caption{The coupled structure is shown schematically. QW and DBR are the quantum
wells, and distributed Bragg reflectors, respectively, while $t_{c}$
is the out-coupling of the cavity. The coupling between the two cavities
is denoted by $\hbar\omega_{12}$.}
\label{fig:coupled-cavity-model-horizontal} 
\end{figure}

The structure that we are going to study is shown schematically in
figure~\ref{fig:coupled-cavity-model-horizontal}; two or three microcavities
are coupled through a partially reflecting mirror. All cavities have
a single quantum well at their centre, which is at the location of
the anti-node of the lowest-lying photon mode of the uncoupled case.
For simplicity, we will not consider cavities with multiple quantum
wells in this paper.

When two or three identical cavities, as described above, are coupled,
the eigenmodes of the photon field are the eigenvectors of the matrices
\[
\left(\begin{array}{cc}
E_{c_{1}} & \hbar\omega_{12}\\
\hbar\omega_{12} & E_{c_{2}}
\end{array}\right)
\]
and 
\[
\left(\begin{array}{ccc}
E_{c_{1}} & \hbar\omega_{12} & 0\\
\hbar\omega_{12} & E_{c_{2}} & \hbar\omega_{23}\\
0 & \hbar\omega_{23} & E_{c_{3}}
\end{array}\right),
\]
respectively. Here the bare cavity energies are denoted by $E_{c_{i}}$,
while $\hbar\omega_{ij}$ is the coupling constant between cavity
$i$, and cavity $j$. In standard structures, $E_{c_{i}}$ is of
the order of a couple of eV, while the coupling is in the meV range.
The eigenvectors are denoted by $\vec{a}^{(i)}$, and, since these
vectors belong to a symmetric matrix (see Appendices 7.2, and 7.3), we have
the orthogonality condition 
\begin{equation}
\vec{a}^{(i)}\cdot\vec{a}^{(j)}=\delta_{ij}.\label{eq:orthogonality}
\end{equation}
For the case of equal cavity energies, and equal couplings, i.e. $\hbar\omega_{12}=\hbar\omega_{23}$,
$\vec{a}^{(i)}$ can be expressed in the double cavity as

\[
\vec{a}^{(1)}=\frac{1}{2}(1,1),\quad\vec{a}^{(2)}=\frac{1}{\sqrt{2}}(1,-1)
\]
and in the triple cavity as 
\begin{align*}
\vec{a}^{(1)} & =\frac{1}{2}(1,\sqrt{2},1),\\
\vec{a}^{(2)} & =\frac{1}{\sqrt{2}}(1,0,-1),\\
\vec{a}^{(3)} & =\frac{1}{2}(1,-\sqrt{2},1).
\end{align*}
The resulting eigenstates are schematically shown in
figure~\ref{fig:photon-symmetry-filled}.
Let us emphasize that, as discussed in the Appendix, the coupling
exciton states have the same symmetry. What is important to note here
is that in the double cavity, the photon states (and consequently,
the polariton states) are either symmetric, or antisymmetric, while
in the triple cavity, the two states that are shifted with respect
to the single cavity are symmetric, while the unshifted state is antisymmetric.
We should keep in mind, however, that the symmetric/antisymmetric
designation of states is not a proper one for non-degenerate cavities,
or unequal coupling constants.

\begin{figure}
\centering
\includegraphics[width=0.65\textwidth]{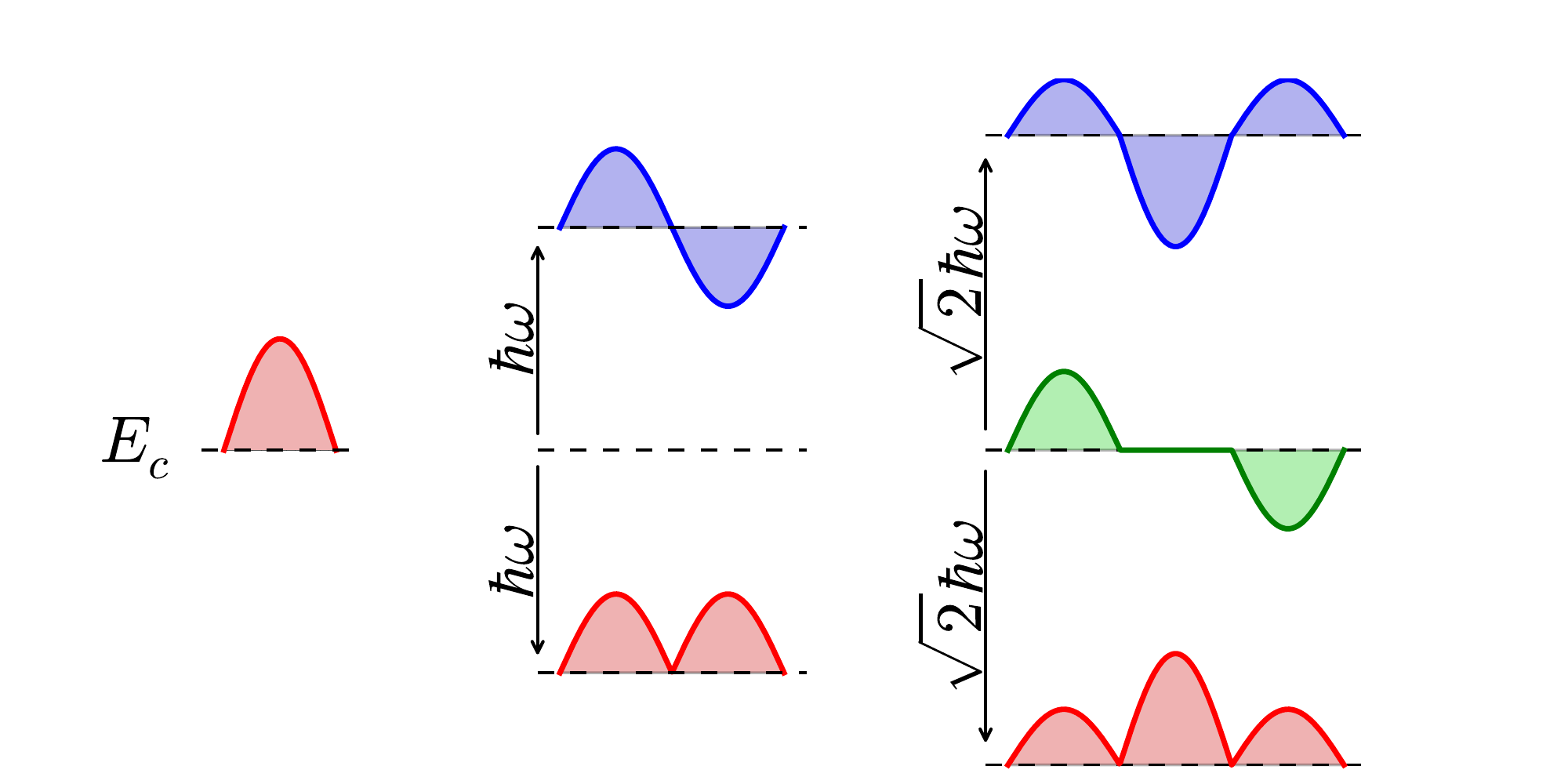}
\caption{Schematic envelope of the electric field in the single (left), double
(centre), and the triple cavity (right). The horizontal axis is the
spatial coordinate. Also shown are the corresponding energy shifts,
when all couplings between the cavities are assumed to be equal to
$\hbar\omega$. \label{fig:photon-symmetry-filled}}
\end{figure}

\begin{figure}
\centering
\includegraphics[width=0.80\textwidth]{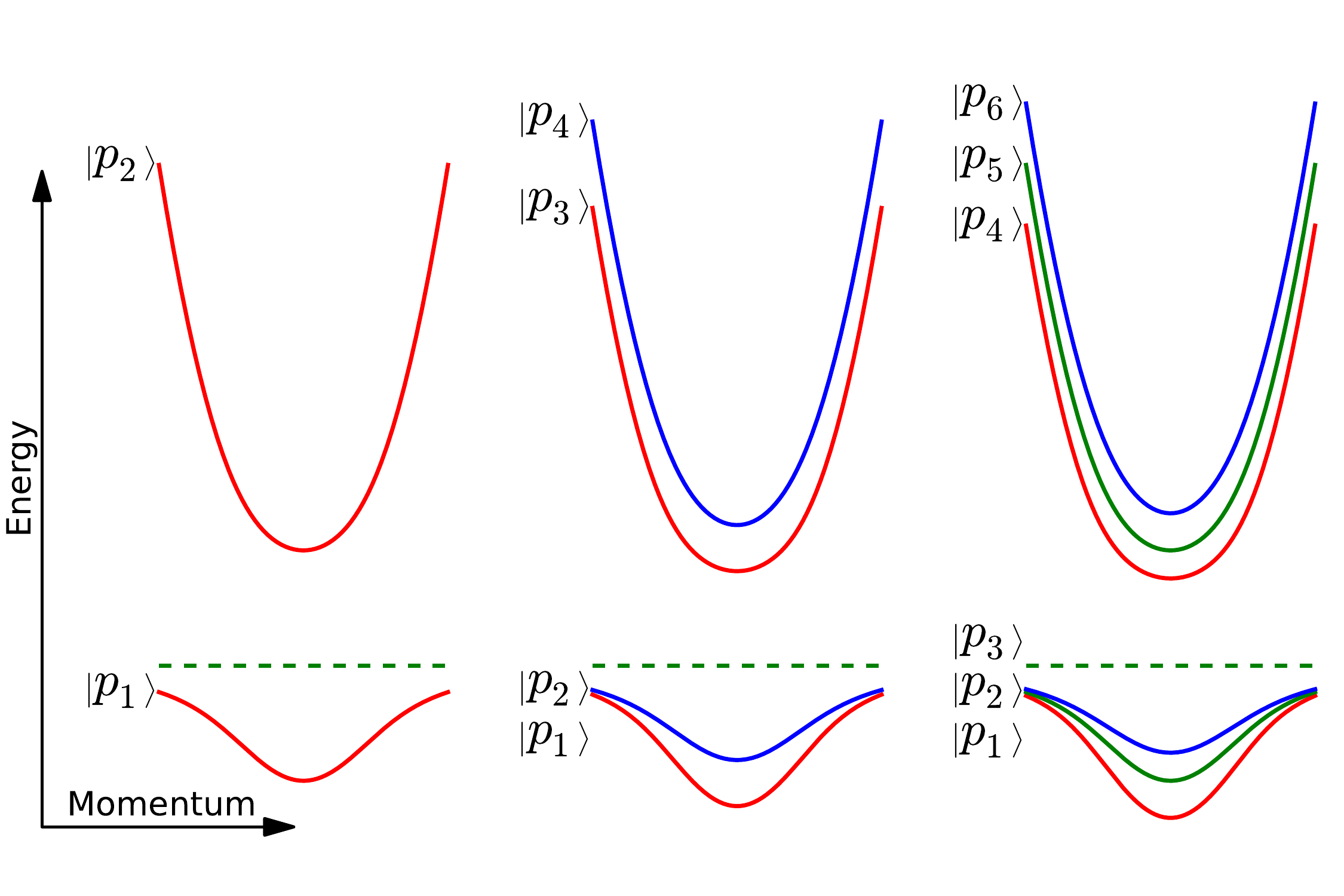}

\caption{The polariton branches (solid lines) with their respective symmetries
in the single (left), double (centre), and triple cavity (right).
Also shown is the exciton reservoir (dashed line). The Rabi splitting,
i.e. the interaction strength between cavity-photons and excitons,
is taken to be $\hbar\Omega_{R}=5\ \mathrm{\mathrm{meV}}$, while
the cavity splittings are $\hbar\omega=2\ \mathrm{meV}$. The colour
of the states is taken from the underlying photon states in
figure~\ref{fig:photon-symmetry-filled}.}
\label{fig:three_polariton_dispersions} 
\end{figure}

As discussed in detail in the Appendix, these photon states, when
coupled to the quantum well excitons, will lead to 2, 4, or 6 polariton
states (branches), depending on how many cavities are coupled. The
dispersion relations of the polariton states are given in
figure~\ref{fig:three_polariton_dispersions}.

\vspace{0pt}

\subsection{Polariton-polariton scattering}

For polariton-polariton processes (i.e. parametric scattering), since
the scattering is mediated by the excitonic part of the particles,
we can separate the intra-cavity and inter-cavity contributions, and
write the transition matrix element as

\begin{eqnarray}
\langle p_{i}p_{j}\vert H_{\mathrm{int}}\vert p_{k}p_{l}\rangle
&=& \sum_{n}a_{n}^{(i)}a_{n}^{(j)} a_{n}^{(k)}a_{n}^{(l)}\langle X_{1}X_{1}\vert
H_{\mathrm{int}}\vert X_{1}X_{1}\rangle
\\
&{}& +
\sum_{\underset{m\neq n}{mn}}a_{n}^{(i)}a_{m}^{(j)}a_{n}^{(k)}a_{m}^{(l)}\langle X_{1}X_{2}
\vert H_{\mathrm{int}}\vert X_{1}X_{2}\rangle.\label{eq:polariton-polariton-matrix-element}
\end{eqnarray}

When contracting the sum, we assumed $\langle X_{1}X_{2}\vert H_{\mathrm{int}}\vert X_{1}X_{2}\rangle
=\langle X_{1}X_{3}\vert H_{\mathrm{int}}\vert X_{1}X_{3}\rangle=\langle X_{2}X_{3}\vert H_{\mathrm{int}}\vert X_{2}X_{3}\rangle$,
i.e., that the coupling Hamiltonian is the same across cavities.

The first term in equation~(\ref{eq:polariton-polariton-matrix-element})
results in a generalized parity condition for the scattering matrix
element, which can be expressed concisely as 
\begin{eqnarray*}
\left(\vec{a}^{(1)}\ast\vec{a}^{(3)}\right)\cdot\left(\vec{a}^{(2)}\ast\vec{a}^{(m)}\right)=0, &  & \qquad m\in\left\{ 1,3\right\} \\
\left(\vec{a}^{(1)}\ast\vec{a}^{(2)}\right)\cdot\left(\vec{a}^{(3)}\ast\vec{a}^{(2)}\right)\neq0,
\end{eqnarray*}
using the pointwise multiplication~$\ast$. Realizing that $\vec{a}^{(1)},\vec{a}^{(3)}$
represent symmetric branches, whereas $\vec{a}^{(2)}$ the anti-symmetric
one, a process $\vert p_{k}p_{l}\rangle\to\vert p_{i}p_{j}\rangle$
(on the lower polariton branches) is allowed, if and only, if 
\begin{equation}
q_{k}+q_{l}=q_{i}+q_{j}\text{ mod }2,\label{eq:parity_triple_cavity}
\end{equation}
where by $q_{k}$ we denote the parity assigned to branch $k$ which,
for the three lower polariton branches, is given by 
\[
q_{1}=0,\qquad q_{2}=1,\qquad q_{3}=0.
\]
For example, two polaritons from the middle branch can scatter one
each to the first and third branches ($q_{2}+q_{2}=q_{1}+q_{3}\text{ mod }2$).
We point out that equation~(\ref{eq:parity_triple_cavity}) holds only
for the case of degenerate cavities ($E_{c_{i}}=E_{c_{j}}$), and
equal couplings ($\hbar\omega_{12}=\hbar\omega_{23}$).

Strictly speaking, we still have to prove that, if this condition
is \textbf{not} satisfied, then the second term of the matrix element
vanishes. However, it is a straightforward computation to show the
desired result, i.e. 
\[
\sum_{\underset{m\neq n}{mn}}a_{n}^{(i)}a_{m}^{(j)}a_{n}^{(k)}a_{m}^{(l)}=0,
\]
follows for all $i,j,k,l$ that violate condition~(\ref{eq:parity_triple_cavity}).
In the double cavity case this condition is just parity conservation,
as we only have one symmetric and one antisymmetric branch both below
and above the exciton reservoir.

\section{Polariton-polariton scattering amplitude in coupled cavities\label{sec:Polariton-polariton-scattering}}

In the preceding section, we showed that polaritons in coupled or
triple cavities fulfil selection rules, when they scatter with either
unlike particles, or other polaritons. However, these rules only tell
us that certain processes are forbidden, but they do not give the
scattering amplitudes. In this section, based on a general framework
for polariton-polariton interaction, we derive the transition probabilities
for polariton-polariton scattering.

Instead of working in the polariton basis, we start by writing down
the equations of motion for the operators of the underlying photon,
and exciton. When doing so, we will assume that both the cavity photons,
and the excitons are confined to their respective cavities, and that
an exciton can interact only with photons in the same cavity. The
exciton-photon couplings are given in the Appendix in
equations~(\ref{eq:single-cavity-hamiltonian},\ref{eq:double-cavity-hamiltonian},\ref{eq:triple-cavity-hamiltonian}).

At least two approaches can be found in the literature for the single
cavity case. First, Ciuti et al. introduced a method based on a Hamiltonian
derived from scattering rates using Fermi's golden rule \cite{Ciuti2003}.
Their approach assumes that polaritons are bosons. Strictly speaking,
this assumption is a valid approximation for low polariton densities
only. Also, polariton decay is only included phenomenologically. On
the other hand, this method contains the symmetries of the system
more explicitly, and one can readily read off, whether a transition
is allowed or not. In addition, it can be augmented to include the
same quantum Langevin based phonon scattering effects, introduced
by Portolan et al. \cite{Portolan2008}.

Second, Portolan et al. developed a method based on the excitonic
equations of motion derived from a microscopic theory of excitons
in a quantum well \cite{Portolan2008c,Portolan2010a}. In this scheme,
a phonon-induced noise term is introduced via the quantum Langevin
approach. No bosonization is used and as such, the scheme is expected
to more closely match experimental data. Thus, in this section, we
will extend the approach by Portolan et al. to the double and triple
cavity case.

In a single cavity, the equations governing the time evolution of
the photon, $a_{\mathbf{k}}$, and exciton, $b_{\mathbf{k}}$, annihilation
operators can be written as (see \cite{Portolan2008c})

\begin{eqnarray}
\frac{d}{dt}a_{\mathbf{k}} & = & -i\left(\omega_{\mathbf{k}}^{c}+it_{c}\right)
a_{\mathbf{k}}+i\Omega_{R}b_{\mathbf{k}}\label{eq:equation-of-motion-photon-single}\\
\frac{d}{dt}b_{\mathbf{k}} & = & -i\omega_{\mathbf{k}}^{x}b_{\mathbf{k}}+
s_{\mathbf{k}}+i\Omega_{R}a_{\mathbf{k}}-\frac{i}{\hbar}R_{\mathbf{k}}^{\mathrm{NL}}
\label{eq:equation-of-motion-exciton-single}
\end{eqnarray}

The term $s_{k}$ only shifts the energy of the
exciton. We did not include excitonic losses, for they are negligible
in comparison with the transmission coefficient $t_{c}$, which denotes
the leakage of photons out of the cavity. Here $\hbar\omega_{\mathbf{k}}^{c}$,
and $\hbar\omega_{\mathbf{k}}^{x}$ are the cavity and exciton energies
at wavevector $\mathbf{k}$, $\hbar\Omega_{R}$ is the Rabi splitting,
and $R_{\mathbf{k}}^{\mathrm{NL}}$ describes the non-linear exciton-exciton
interaction, and can be expressed as

\begin{align*}
R_{\mathbf{k}}^{\mathrm{NL}} & =R_{\mathbf{k}}^{\mathrm{xx}}+R_{\mathbf{k}}^{\mathrm{sat}}\\
R_{\mathbf{k}}^{\mathrm{xx}} & =V_{\mathrm{xx}}\sum_{\mathbf{k}_{1}\mathbf{k}_{2}}
b_{\mathbf{k}_{1}+\mathbf{k}_{2}-\mathbf{k}}^{\dagger}b_{\mathbf{k}_{1}}^{\phantom{\dagger}}
b_{\mathbf{k}_{2}}^{\phantom{\dagger}}\\
R_{\mathbf{k}}^{\mathrm{sat}} & =\frac{V}{n_{\mathrm{sat}}}\sum_{\mathbf{k}_{1}\mathbf{k}_{2}}
b_{\mathbf{k}_{1}+\mathbf{k}_{2}-\mathbf{k}}^{\dagger}b_{\mathbf{k}_{1}}^{\phantom{\dagger}}
a_{\mathbf{k}_{2}}^{\phantom{\dagger}}.
\end{align*}
\modification{In the equations above, $V_{\mathrm{xx}}$ is the exciton-exciton interaction potential, which can be
approximated as a momentum-independent constant, $V_{xx}=6e^{2}/(\pi\epsilon\lambda_{x})$, where $\lambda_x$ is
the exciton Bohr radius, $\epsilon$ is the dielectric contant, and $e$ is the electron charge. $n_{\mathrm{sat}}$
denotes the exciton saturation density with the value $n_{\mathrm{sat}} = 7/(16\pi\lambda_x^2)$ \cite{Ciuti2003}.}
Using equation~(\ref{eq:transformation-1}), and denoting the polariton
annihilation operator by $P_{i\bld k}$, we can write down the equations
of motion in the polariton basis, and they take on the form

\begin{align}
\frac{d}{dt}P_{1\bld k} & =-i\omega_{1\bld k}P{}_{1\bld k}+\tilde{E}_{1\bld k}^{\mathrm{in}}
-i\tilde{R}_{1\bld k}^{\mathrm{NL}}\label{eq:portolan_polariton_1}\\
\frac{d}{dt}P_{2\bld k} & =-i\omega_{2\bld k}P_{2\bld k}+\tilde{E}_{2\bld k}^{\mathrm{in}}
-i\tilde{R}_{2\bld k}^{\mathrm{NL}},\label{eq:portolan_polariton_2}
\end{align}
with

\[
\tilde{R}_{i\bld k}^{\mathrm{NL}}=c_{i\bld k}R_{\bld k}^{\mathrm{NL}},
\qquad\tilde{E}_{i\bld k}^{\mathrm{in}}=(-1)^{i}c_{3-i\bld k}t_{c}E_{\bld k}^{\mathrm{in}}.
\]

The interaction term, $R_{\mathbf{k}}^{\mathrm{NL}}$, in the polariton
basis is given by

\begin{align*}
R_{\mathbf{k}}^{\mathrm{xx}} & =-V_{\mathrm{xx}}\sum_{\bld k_{1}
\bld k_{2}jj_{1}j_{2}}(-1)^{j+j_{1}+j_{2}}c_{j\bld k_{1}+\bld k_{2}-\bld k}c_{j_{1}\bld k_{1}}
c_{j_{2}\bld k_{2}}p_{j\bld k_{1}+\bld k_{2}-\bld k}^{\dagger}p_{j_{1}\bld k_{1}}p_{j_{2}\bld k_{2}}\\
R_{\bld k}^{\mathrm{sat}} & =\frac{V}{n_{\mathrm{sat}}}\sum_{\bld k_{1}\bld k_{2}jj_{1}j_{2}}
(-1)^{j+j_{1}}c_{j\bld k_{1}+\bld k_{2}-\bld k}c_{j_{1}\bld k_{1}}c_{3-j_{2}\bld k_{2}}
p_{j\bld k_{1}+\bld k_{2}-\bld k}^{\dagger}p_{j_{1}\bld k_{1}}p_{j_{2}\bld k_{2}}
\end{align*}
where the branch indices are denoted by $j,j_{1},j_{2}$.

The equations of motion in the general case can readily be written
down, if we notice that a photon and an exciton are coupled only,
if they are located in the same cavity, and that photons in adjacent
cavities are coupled by tunnelling. We denote the energy associated
with cavity-cavity coupling by $\hbar\omega_{jl}$. With this extra
term, equations~(\ref{eq:equation-of-motion-photon-single}-\ref{eq:equation-of-motion-exciton-single})
become

\begin{eqnarray}
\frac{d}{dt}a_{\mathbf{k}}^{j} & = & -i\left(\omega_{\mathbf{k}}^{c_{j}}+it_{j}\right)
a_{\mathbf{k}}^{j}+i\sum_{l\ne j}\omega_{jl}a_{\mathbf{k}}^{l}+i\Omega_{j}
b_{\mathbf{k}}^{j}\label{eq:equation-of-motion-photon-general}\\
\frac{d}{dt}b_{\mathbf{k}}^{j} & = & -i\omega_{\mathbf{k}}^{x{}_{j}}
b_{\mathbf{k}}^{j}+s_{\mathbf{k}}+i\Omega_{j}a_{\mathbf{k}}^{j}-
\frac{i}{\hbar}R_{\mathbf{k}}^{\mathrm{NL},j}\ .
\label{eq:equation-of-motion-exciton-general}
\end{eqnarray}
We can then derive the appropriate expressions for $R_{\bld k}^{\mathrm{xx},i}$
and $R_{\boldsymbol{k}}^{\mathrm{sat},i}$, where $i$ denotes the
cavity index. For the sake of brevity we give the result, for the
double cavity case, in Appendix \ref{sub:interaction-terms}.

Similar (but more complicated) expressions hold for the case of the
triple cavity. Their derivation is straightforward, but quite tedious,
thus, we skip it here.

Once we have the equations of motion in the polariton basis, we can
study any parametric scattering scheme by simply fixing the wave vectors,
and the branch indices of the pumps, the signal, and the idler. These
results are used to derive the coupling coefficients in the next section.

\section{Parametric scattering schemes}

Before presenting the numerical results, it is instructive to investigate the possible parametric processes in a single,
double or triple cavity. \modification{In what follows, in order to simplify references to the various scattering
schemes, we introduce the notation \scheme{ij}{kl}{s,d,t}, where $ij$ stand for the branch indices of the two pump
polaritons, while $kl$ are the branch indices of the signal-idler pair polaritons. Finally, $s, d$, and $t$ designate
the cavity configuration, i.e., whether we are dealing with a $s$ingle, $d$ouble, or $t$riple cavity. For example,
\scheme{22}{12}{s} would denote the single-cavity process that takes two pump polaritons from the second (upper)
polaritons branch to the first (lower) and second polariton branches. This process would correspond to the scheme
proposed by C. Ciuti \cite{Ciuti2004}.} We begin our discussion with the case of the single cavity.

\subsection{Parametric scattering in a single cavity}

In this case, we have two polariton branches, and two input polaritons
that we have to distribute on them. The branch-entanglement scheme
of Ciuti \cite{Ciuti2004}, or any other scheme that has at least
one pump on the upper branch, suffers from the above-mentioned problem
of signal polaritons' leaking to the exciton reservoir. This can only
be avoided, if both pump polaritons are on the lowest branch. Such
a scheme was proposed by Portolan et al. \cite{Portolan2009}. We
will denote this scheme by \modification{\scheme{11}{11}{s}}. It has already been published in the
cited work, therefore, we include it only for the sake of comparison
with other schemes.

\subsection{Parametric scattering in a double cavity}

\begin{figure}
\centering
\includegraphics[width=0.49\columnwidth]{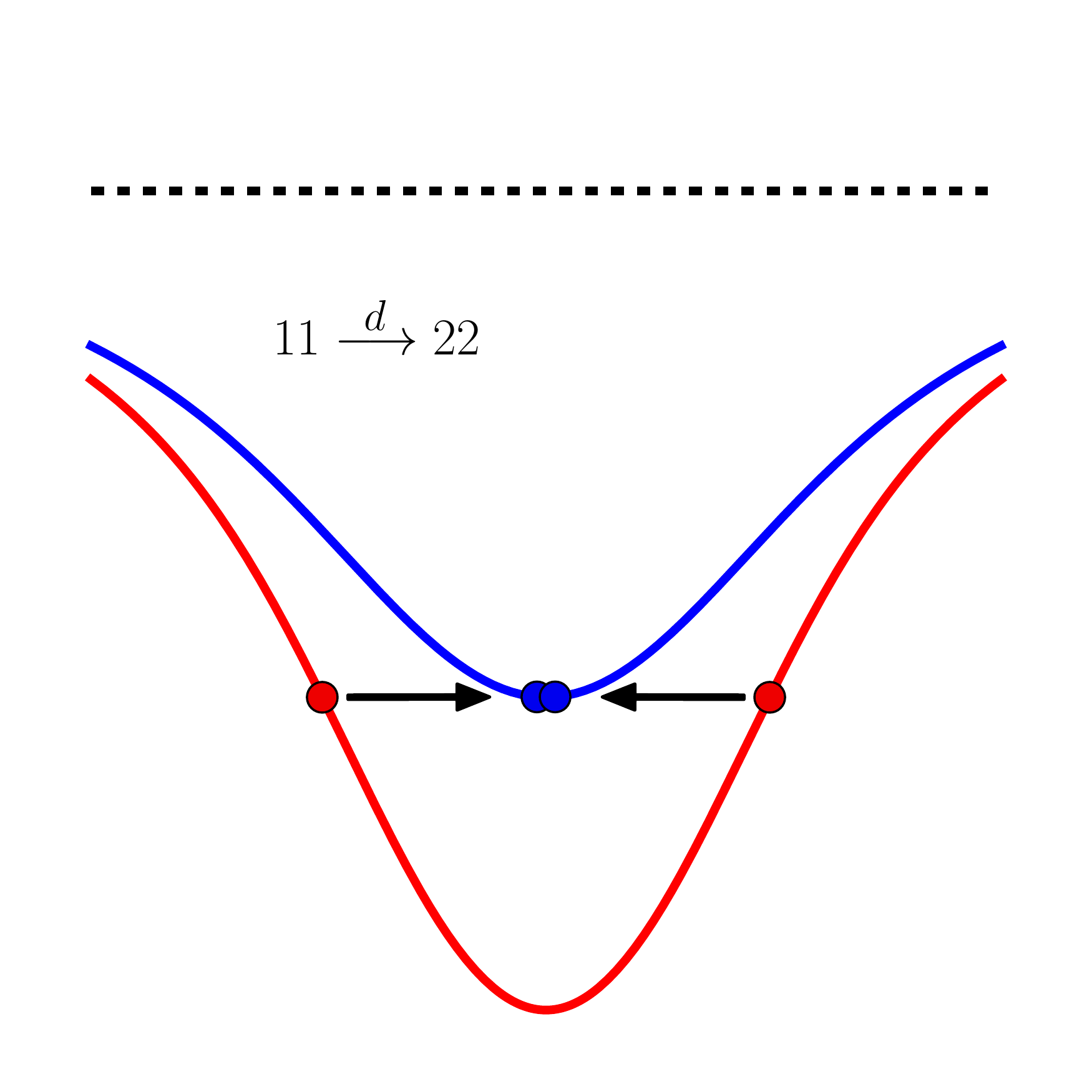}
\includegraphics[width=0.49\columnwidth]{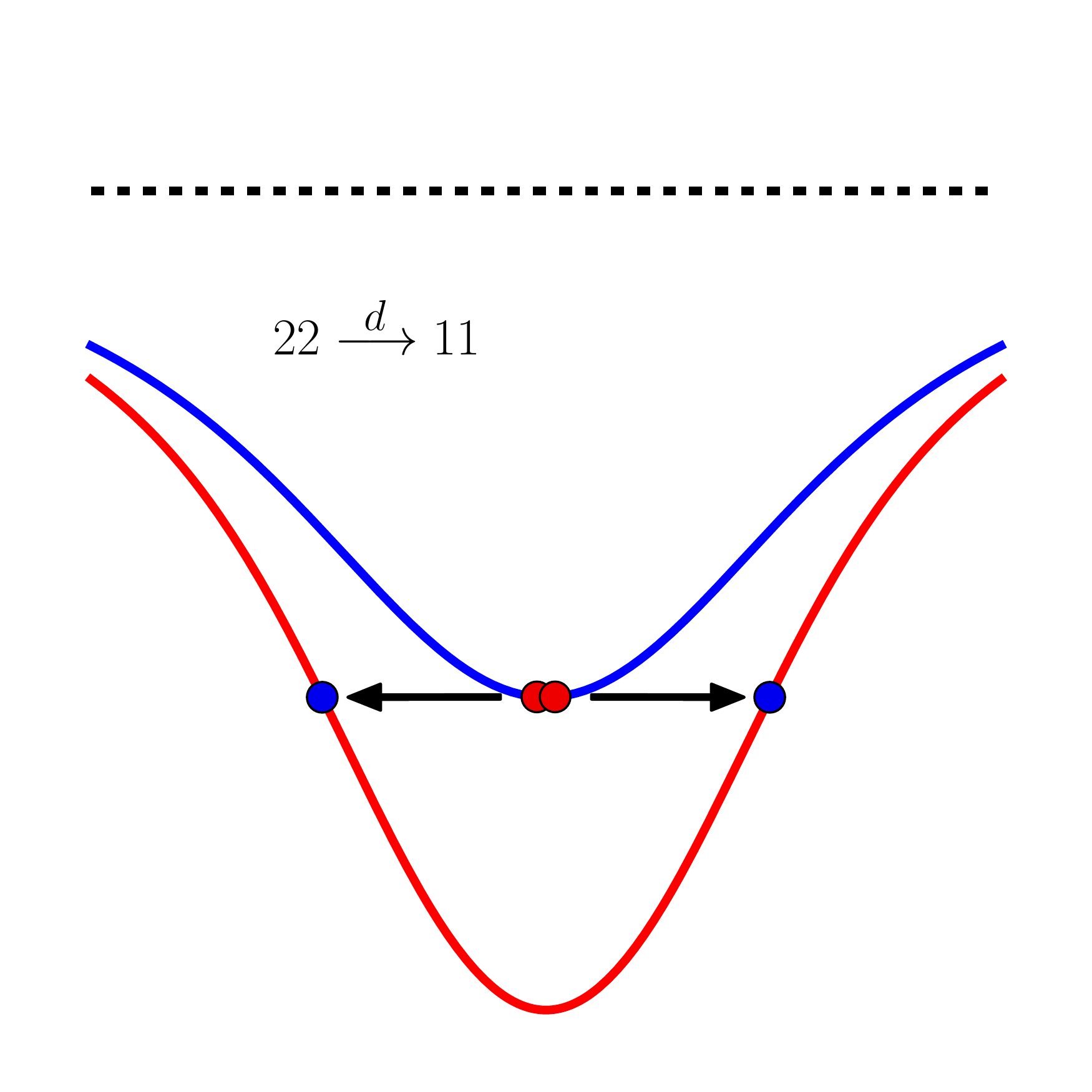}

\includegraphics[width=0.49\columnwidth]{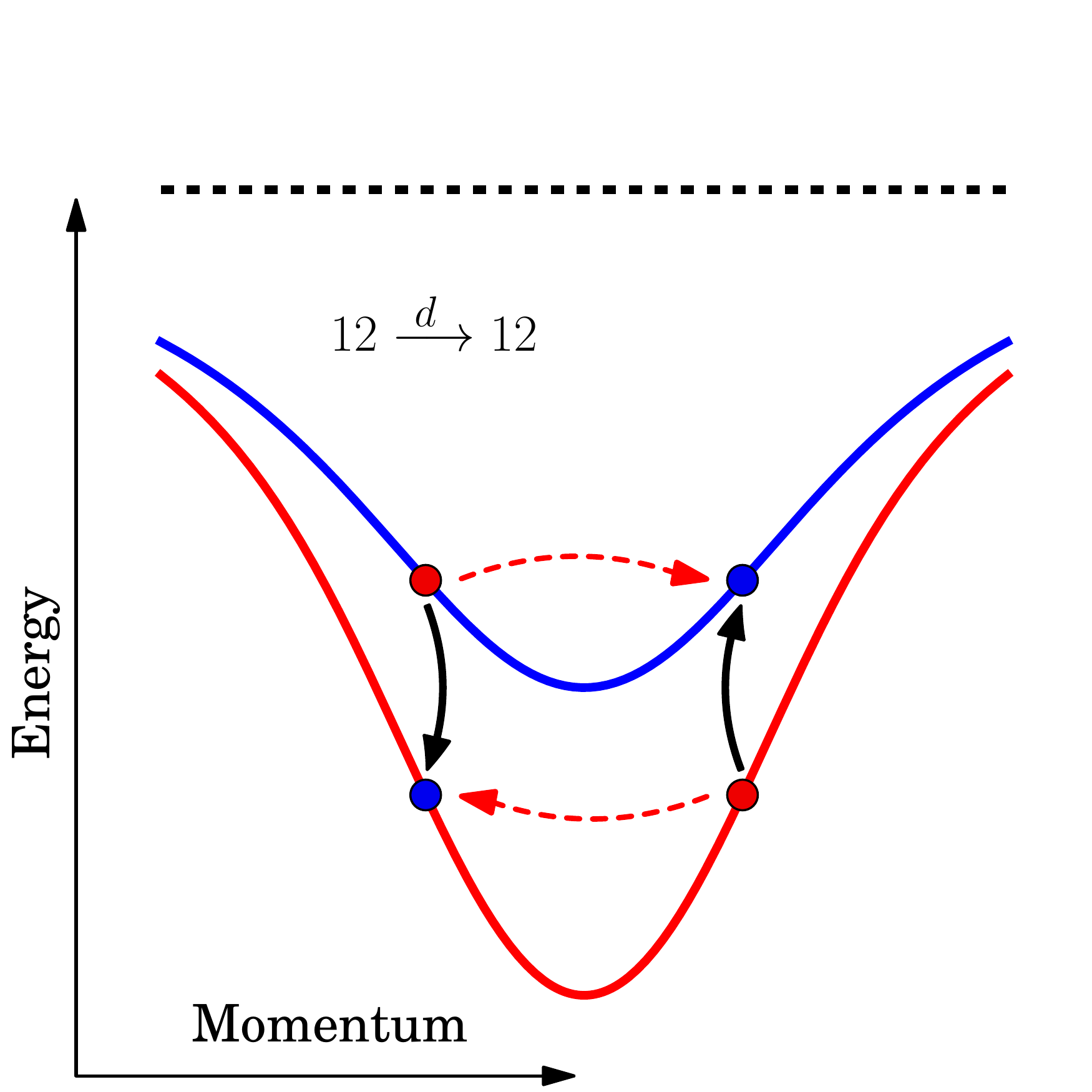}
\includegraphics[width=0.49\columnwidth]{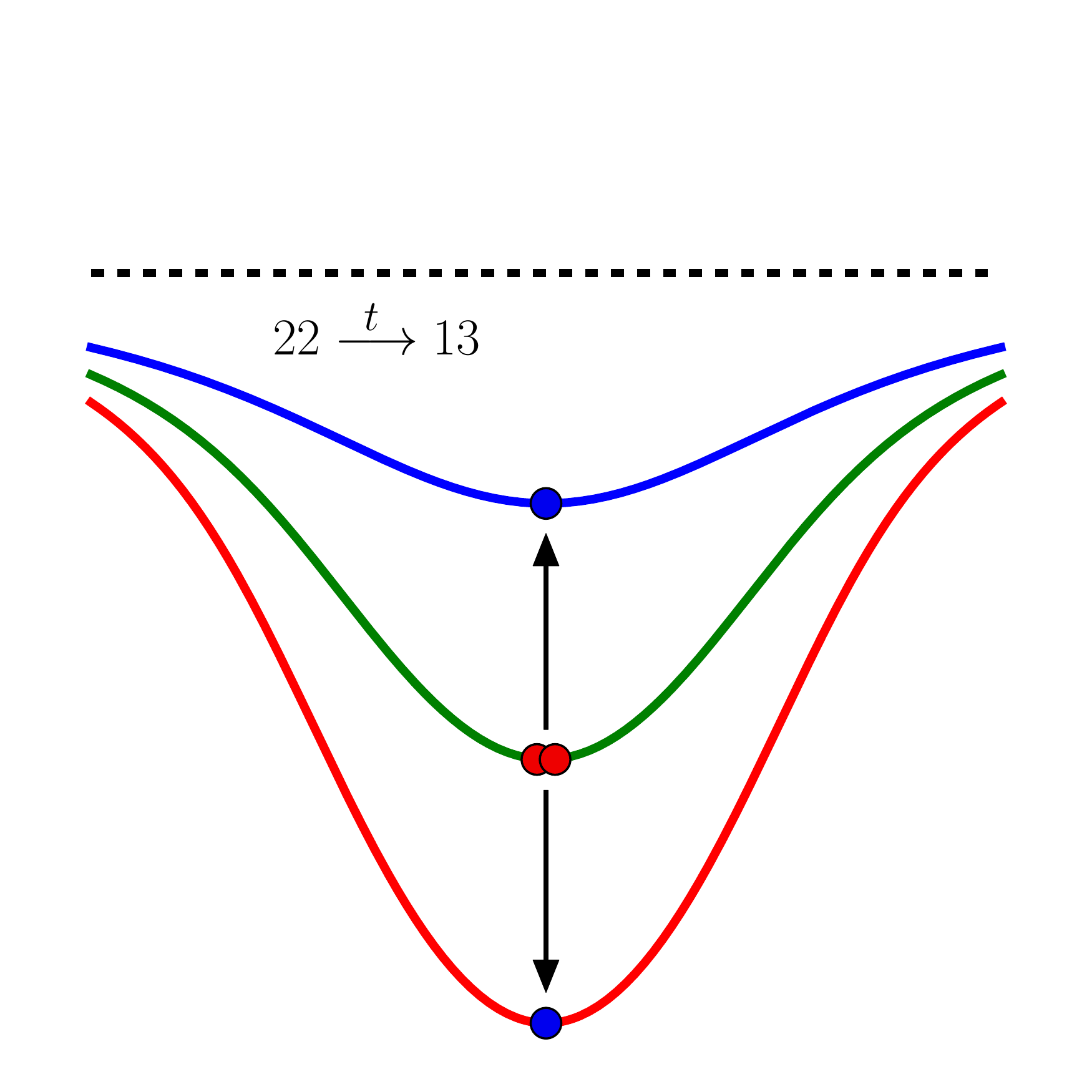}

\caption{Allowed parametric scattering in a double cavity. Pump polaritons
are denoted by red dots, while the signal-idler pair by blue dots.
The dashed line is the exciton dispersion. \modification{The designation of the scattering processes is explained in the
text}.}
\label{fig:double-cavity-allowed} 
\end{figure}

As we have already pointed out, we are interested in scattering on
the lower branches. Therefore, we have to distribute the two input,
and two output polaritons on the two branches in such a way that both
energy and momentum are conserved, and the parity conservation rules
discussed in Section \ref{sec:symmetry} are satisfied. Keeping these
constraints in mind, the processes shown in
figure~\ref{fig:double-cavity-allowed}
are allowed. We note here that the entanglement-generating
scheme of Ciuti \cite{Ciuti2004}, in which the two pump polaritons
are at $k=0$, and the signal-idler pair is on two different branches,
cannot be realized on the two lower polariton branches, because it
would violate the the parity conservation rule. We do not include
the intra-branch magic angle scattering \cite{Savvidis2000a, Savvidis2000}, and the scheme similar to
\modification{\scheme{11}{11}{s}} for the following reasons: in the magic angle scattering, the
signal and idler are at different energies, and this leads to both
distinguishability, and highly different decay times, therefore, it
is not a good candidate for entanglement generation, while the original
scheme of Portolan et al. do not qualitatively differ in the single
or double cavity cases.

However, due to the modified structure of the lower
polariton branches we can implement the scattering processes illustrated
in figure~\ref{fig:double-cavity-allowed}.

\subsection{Parametric scattering in a triple cavity}

The case of the triple cavity is similar to that of the double cavity,
i.e., for any two of the three lower polariton branches, \modification{\scheme{11}{22}{d}, \scheme{22}{11}{d},
and \scheme{12}{12}{d}} in figure~\ref{fig:double-cavity-allowed} would still be allowed,
but in addition, we would also have the possibility of two polaritons
scattering from the second branch to the first and third branch, \modification{\scheme{22}{13}{t},} or
the reverse process, \modification{(\scheme{13}{22}{t}),} as shown in figure~\ref{fig:double-cavity-allowed}.
Due to energy conservation, this latter process
has to be vertical, if $\hbar\omega_{12}=\hbar\omega_{23}$, i.e.,
both the pump polaritons and the signal-idler pair are located at
$k=0.$

\section{Analytical modelling\label{sec:analytical-modeling}}

In order to develop the intuition for the dynamics in polariton scattering,
we start by considering a very simple model characterized only by
three parameters, the decay width $\Gamma$ \modification{(which is mainly determined by the excitonic and photonic
linewidths)}, the dimensionless noise
background $n$ (\modification{In this simple model the background noise includes detector noise, pump-induced
photoluminescence due to the finite temperature, etc.)}, and $\tilde{\Delta}$, which combines the pump and coupling
strength. For the noise is an external parameter, it is clear that
this model will not be able to distinguish between the merits of different
pump schemes. We will tackle this question in Section \ref{sec:numerical-simulations}.

\modification{Once the signal-idler states are fixed, the general equations of motion
(\ref{eq:portolan_polariton_1},\ref{eq:portolan_polariton_2}) decouple with respect to the momentum degree of freedom
and yield a pair of two differential equations, one for the signal, $P_s$, and the other for the idler, $P^\dagger_i$
\cite{Portolan2008c}. The coupled two-mode system
of equation reads}
\begin{eqnarray}
\frac{d}{dt}P_s &=& - i \left(\omega_s-i\frac{\Gamma_s}{2}\right)P_s-ig \mathcal{P}_1\mathcal{P}_2 P^\dagger_i  +
\mathcal{F}_s, 
\\
\frac{d}{dt}P^\dag_i &=& i\left( \omega_i  +i \frac{\Gamma_i}{2} \right) P^\dagger_i +i g^*
\mathcal{P}_1^* \mathcal{P}_2^* P_{s} + \mathcal{F}^\dagger_i\ ,
\end{eqnarray}
\modification{where $\mathcal{P}_{1,2}$ are the expectation values corresponding to the pump 
operators $P_{1,2}$, and the background noise $n$ enters through the time-dependent Langevin noise operators
$\mathcal{F}_s$ and $\mathcal{F}_i$. 
Under continuous driving conditions the effect of noise is fully determined by the stationary correlators $\langle
\mathcal{F}^\dagger_m(t) \mathcal{F}_m(t') \rangle = \Gamma n \delta(t-t')$ and  $\langle \mathcal{F}_m(t) 
\mathcal{F}^\dagger_m(t') \rangle = \Gamma (n+1)\delta(t-t') $, where $\Gamma$ is the total polariton decay
rate of the particular mode, and $m=i,s$. Note that $n$ is a function of the temperature, as well as of the cavity
under consideration. This parameter can be determined by solving a Boltzmann-type equation \cite{Portolan2008c}.}

The equations of motion of this model can be solved analytically in
the steady state \cite{Portolan2008}, and the signal population $N_{s}(t)$
can be expressed as

\begin{align*}
N_{s}(t) & =\vert c_{1}^{s}(0,t)\vert^{2}N_{s}(0)+\vert c_{2}^{s}(0,t)\vert^{2}\left(N_{i}(0)+1\right)\\
 & \qquad+\Gamma n\int_{0}^{t}d\tau\,\vert
c_{1}^{s}(\tau,t)\vert^{2}+\Gamma\left(n+1\right)\int_{0}^{t}d\tau\,\vert
c_{2}^{s}(\tau,t)\vert^{2}
\end{align*}
while the correlators take on the form

\begin{align*}
\langle P_{s}^{\dagger}(t_{1})P_{i}^{\dagger}(t_{2})\rangle & =c_{1}^{s}(0,t_{1})^{*}
c_{2}^{i}(0,t_{2})N_{s}(0)+c_{2}^{s}(0,t_{1})^{*}c_{1}^{i}(0,t_{2})\left(N_{i}(0)+1\right)\\
 & \qquad+\Gamma n\int_{0}^{\text{min}(t_{1},t_{2})}du\, c_{1}^{s}(u,t_{1})c_{2}^{i}
(u,t_{2})^{*}\\
 & \qquad +\Gamma(n+1)\int_{0}^{\text{min}(t_{1},t_{2})}du\,
c_{2}^{s}(u,t_{1})c_{1}^{i}(u,t_{2})^{*},
\end{align*}
where

\begin{align*}
c^{s,i}_{1}(t_{1},t_{2}) &
=e^{\left(-i\omega_{s,i}-\frac{\Gamma}{2}\right)(t_{2}-t_{1})}\cosh\left(\tilde{\Delta}(t_{2}-t_{1})\right)\\
c^{s,i}_{2}(t_{1},t_{2}) &
=e^{\left(-i\omega_{s,i}-\frac{\Gamma}{2}\right)(t_{2}-t_{1})}\sinh\left(\tilde{\Delta}(t_{2}-t_{1})\right).
\end{align*}

The pump, as well as, the coupling strength enter these equations
via the $\tilde{\Delta}$ parameter which is given by 
\[
\tilde{\Delta}=g\mathcal{P}_{1}\mathcal{P}_{2},
\]
where $\mathcal{P}_{1},\mathcal{P}{}_{2}$ are constant in time. 

Since all the information of the present model is contained in the three parameters $\Gamma$, $n$ and $\tilde{\Delta}$,
we can apply it equally well to the single and double cavity case. We will see in the subsequent discussion
that the double cavity has some advantages over the single cavity, if phonon-induced photoluminescence is considered.

We proceed by writing out the explicit expression for the population 
and correlation in the steady state (assuming that $\Gamma > 2\tilde\Delta$) 
\begin{align*}
N_{s}(t\to\infty)=N_{i}(t\to\infty) & =\frac{n\Gamma^{2}+2\tilde{\Delta}^{2}}{\Gamma^{2}-4\tilde{\Delta}^{2}}\\
\langle P_{s}^{\dagger}(t\to\infty)P_{i}^{\dagger}(t\to\infty)\rangle\langle P_{s}(t\to\infty)
P_{i}(t\to\infty)\rangle & =\left(\frac{(1+2n)\Gamma\tilde{\Delta}}{\Gamma^{2}-4\tilde{\Delta}^{2}}\right)^2.
\end{align*}

\modification{In order to tomographically reconstruct the populations and correlations, we have to write down the
two-particle density matrix \cite{James2001}.} By denoting the relative phase of the two pump beams by $\Theta$,
after some simple, but tedious algebra, the density matrix can be written as (we have renormalized $\tilde{\Delta}$ by
$\Gamma$ such that the two dimensionless quantities $\Delta=\tilde{\Delta}/\Gamma$ and $n$ remain as the only
parameters)

\begin{align}
\rho & =\frac{1}{16\Delta^{4}+2(4n(n+3)+1)\Delta^{2}+4n^{2}}\\
 & \quad\cdot\left[\begin{array}{cccc}
\rho_{11} & 0 & 0 & e^{-4i\theta}(2n\Delta+\Delta)^{2}\\
0 & \left(2\Delta^{2}+n\right)^{2} & 0 & 0\\
0 & 0 & \left(2\Delta^{2}+n\right)^{2} & 0\\
e^{4i\theta}(2n\Delta+\Delta)^{2} & 0 & 0 & \rho_{11}
\end{array}\right],
\label{eq:density-matrix-steady-state}
\end{align}
where $\rho_{11} = 4\Delta^{4}+(4n(n+2)+1)\Delta^{2}+n^{2}$. This experssion for the density matrix allows us to
calculate the entanglement of formation (EOF) with respect to the polarization degree of freedom as a function of both
the (uniform) noise background, and the pump intensities \cite{Wootters1998,Hill1997}. \modification{The EOF has a
direct operational meaning as the minimum amount of information needed to form  the entangled state under investigation
out of uncorrelated ones.}

The result is plotted in
figure~\ref{fig:eof_steady_state}, while an approximate, but
physically more transparent expression in given in the Appendix, in
\ref{sub:approx-eof}.

\begin{figure}
\centering
\includegraphics[width=0.98\textwidth]{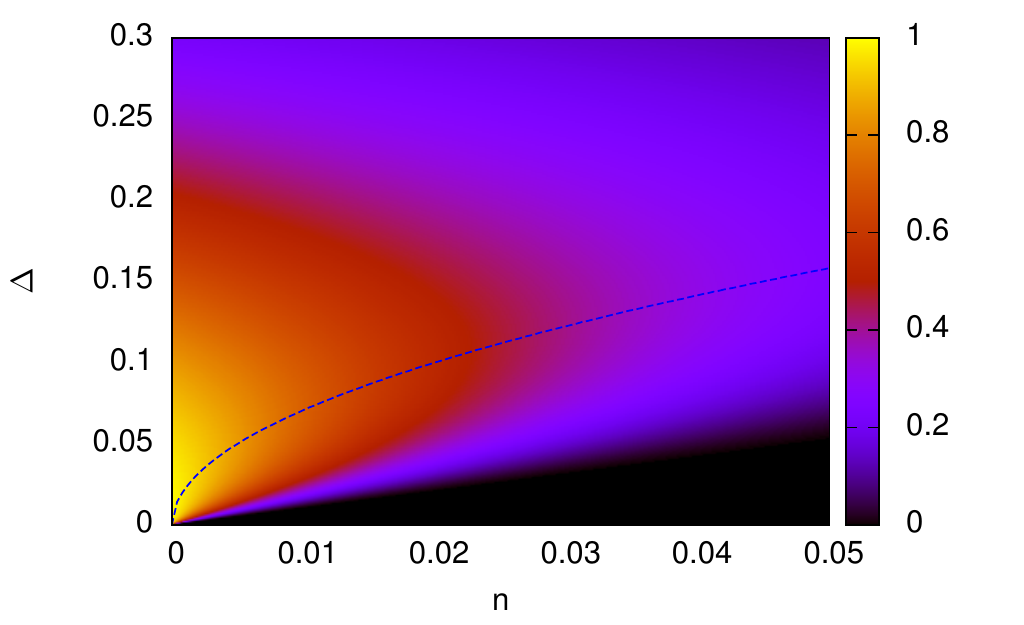}
\caption{The entanglement of formation as a function of $\Delta$ and the noise
background $n$. Also shown is the contour of the maximum achievable
entanglement. The black region ($\Delta<n$) corresponds to zero entanglement
of formation.}
\label{fig:eof_steady_state} 
\end{figure}

It is no surprise that the higher the noise background, the harder
we have to pump, and that the maximal achievable entanglement of formation
decreases with increasing noise background. It should, however, be
noted that even though photoluminescence (and thus temperature effects)
can be included in this model in the noise background $n$, this is
insufficient to evaluate the merit of different pump schemes, as we
are not able to ascertain the dependence of $n$ on the temperature.
In any instance, the maximum achievable entanglement is already significantly
reduced not only for moderate values of $n$ but also for moderate
values of the pump intensity. In addition, the pump intensity needed
to achieve the maximal entanglement of formation for a fixed noise
background is not a linear function of $n$.

For a realistic system, temperature effects and pump induced photoluminescence,
in addition, decrease the achievable entanglement. This is the topic
of the next section.

\section{Numerical simulations\label{sec:numerical-simulations}}

In the previous section, we saw that an analytical model cannot capture
all details of the polariton-polariton scattering problem at hand.
For a fuller understanding, we have to resort to the numerical solutions
of equations~(\ref{eq:equation-of-motion-photon-general}-\ref{eq:equation-of-motion-exciton-general}).

By applying the procedure outlined in section \ref{sec:Polariton-polariton-scattering}
(i.e. fixing the signal and idler as well as the pump configuration)
and promoting the two remaining equations to the desired Heisenberg-Langevin
equations, we get 
\begin{align}
\frac{d}{dt}P_{j_{s}\bld k_{s}} & =-i\tilde{\omega}_{\bld k}P_{j_{s}\bld k_{s}}+
g_{\bld k}\mathcal{P}_{\bld k_{p_{1}}}\mathcal{P}_{\bld k_{p_{2}}}P_{j_{i}\bld k_{i}}+\mathcal{F}_{P_{s}}
\label{eq:langevin_signal}\\
\frac{d}{dt}P_{j_{i}\bld k_{i}}^{\dagger} & =i\tilde{\omega}_{\bld k_{i}}P_{j_{i}\bld k_{i}}
+g_{\bld k}^{*}\mathcal{P}^{*}_{\bld k_{p_{1}}}\mathcal{P}^{*}_{\bld k_{p_{2}}}P_{j_{s}\bld
k_{s}}+\mathcal{F}_{P_{i}}^{\dagger}.
\label{eq:langevin_idler}
\end{align}
Here $\mathcal{F}_{P_{s,i}}$ are Markovian noise operators \cite{Portolan2008}.

We duly note that these equations are the same for Portolan's as well
as Ciuti's approach. The difference manifests itself only in a different
value for the coupling $g_{\mathbf{k}}$, as shown in Table~\ref{tab:g_k_summary}.

\begin{table}
\centering
\rotatebox{90}{
\begin{tabular}{ccccccc}
\hline 
{\small Scheme}  &  & Model  &  & coefficient of $2V_{\mathrm{xx}}$  &  & coefficient of $\frac{V}{n_{\mathrm{sat}}}$\tabularnewline
\hline 
 &  &  &  &  &  & \tabularnewline
\modification{\scheme{11}{11}{s}} &  & Portolan  &  & $c_{1\bld k}c_{1\bld k_{i}}c_{1\bld k_{p_{1}}}c_{1\bld
k_{p_{2}}}$  &  & $c_{1\bld k}c_{1\bld k_{i}}\left[c_{1\bld k_{p_{1}}}c_{2\bld k_{p_{2}}}+c_{2\bld k_{p_{1}}}c_{1\bld
k_{p_{2}}}\right]$\tabularnewline
 &  & Ciuti  &  & $c_{1\bld k}c_{1\bld k_{i}}c_{1\bld k_{p_{1}}}c_{1\bld k_{p_{2}}}$  &  & $2\left[c_{2\bld k_{i}}c_{1\bld k}+c_{2\bld k}c_{1\bld k_{i}}\right]c_{1\bld k_{p_{1}}}c_{1\bld k_{p_{2}}}+2c_{1\bld k}c_{1\bld k_{i}}\left[c_{1\bld k_{p_{1}}}c_{2\bld k_{p_{2}}}+c_{2\bld k_{p_{1}}}c_{1\bld k_{p_{2}}}\right]$\tabularnewline
 &  &  &  &  &  & \tabularnewline
\modification{\scheme{11}{11}{d}} &  & Portolan  &  & $\frac{1}{2}c_{1\bld k}c_{1\bld k_{i}}c_{1\bld k_{p_{1}}}c_{1\bld
k_{p_{2}}}$  &  & $\frac{1}{2}c_{1\bld k}c_{1\bld k_{i}}\left[c_{1\bld k_{p_{1}}}c_{2\bld k_{p_{2}}}+c_{1\bld
k_{p_{2}}}c_{2\bld k_{p_{1}}}\right]$\tabularnewline
 &  & Ciuti  &  & $\frac{1}{2}c_{1\bld k}c_{1\bld k_{i}}c_{1\bld k_{p_{1}}}c_{1\bld k_{p_{2}}}$  &  & $c_{1\bld k_{p_{1}}}c_{2\bld k_{p_{2}}}\left[c_{1\bld k}c_{2\bld k_{i}}+c_{2\bld k}c_{1\bld k_{i}}\right]+\left[c_{1\bld k_{p_{1}}}c_{2\bld k_{p_{2}}}+c_{2\bld k_{p_{1}}}c_{1\bld k_{p_{2}}}\right]c_{1\bld k}c_{1\bld k_{i}}$\tabularnewline
 &  &  &  &  &  & \tabularnewline
\modification{\scheme{11}{22}{d}} &  & Portolan  &  &
$\frac{1}{2}c_{3\boldsymbol{k}}c_{3\boldsymbol{k}_{i}}c_{1\boldsymbol{k}_{p1}}c_{1\boldsymbol{k}_{p2}}$  &  &
$\frac{1}{2}c_{3\boldsymbol{k}}c_{3\boldsymbol{k}_{i}}\left[c_{1\boldsymbol{k}_{p1}}c_{1\boldsymbol{k}_{p2}}+c_{
1\boldsymbol{k}_{p2}}c_{2\boldsymbol{k}_{p1}}\right]$\tabularnewline
 &  & Ciuti  &  & $\frac{1}{2}c_{3\bld k}c_{3\bld k_{i}}c_{1\bld k_{p_{1}}}c_{1\bld k_{p_{2}}}$  &  & $c_{1\bld k_{p_{1}}}c_{1\bld k_{p_{2}}}\left[c_{3\bld k}c_{4\bld k_{i}}+c_{4\bld k}c_{3\bld k_{i}}\right]+\left[c_{1\bld k_{p_{1}}}c_{2\bld k_{p_{2}}}+c_{2\bld k_{p_{1}}}c_{1\bld k_{p_{2}}}\right]c_{3\bld k}c_{3\bld k_{i}}$\tabularnewline
 &  &  &  &  &  & \tabularnewline
\modification{\scheme{22}{11}{d}} &  & Portolan  &  &
$\frac{1}{2}c_{1\boldsymbol{k}}c_{1\boldsymbol{k}_{i}}c_{3\boldsymbol{k}_{p1}}c_{3\boldsymbol{k}_{p2}}$  &  &
$\frac{1}{2}c_{1\boldsymbol{k}}c_{1\boldsymbol{k}_{i}}\left[c_{3\boldsymbol{k}_{p1}}c_{4\boldsymbol{k}_{p2}}+c_{
3\boldsymbol{k}_{p2}}c_{4\boldsymbol{k}_{p1}}\right]$\tabularnewline
 &  & Ciuti  &  & $\frac{1}{2}c_{1\bld k}c_{1\bld k_{i}}c_{3\bld k_{p_{1}}}c_{3\bld k_{p_{2}}}$  &  & $c_{3\bld k_{p_{1}}}c_{3\bld k_{p_{2}}}\left[c_{1\bld k}c_{2\bld k_{i}}+c_{2\bld k}c_{1\bld k_{i}}\right]+\left[c_{3\bld k_{p_{1}}}c_{4\bld k_{p_{2}}}+c_{4\bld k_{p_{1}}}c_{3\bld k_{p_{2}}}\right]c_{1\bld k}c_{1\bld k_{i}}$\tabularnewline
\end{tabular}
}
\caption{Coupling coefficient $g_{\bld k}$ for different schemes with respect
to both models.\label{tab:g_k_summary}}
\end{table}

\begin{figure}
\includegraphics[width=0.7\columnwidth]{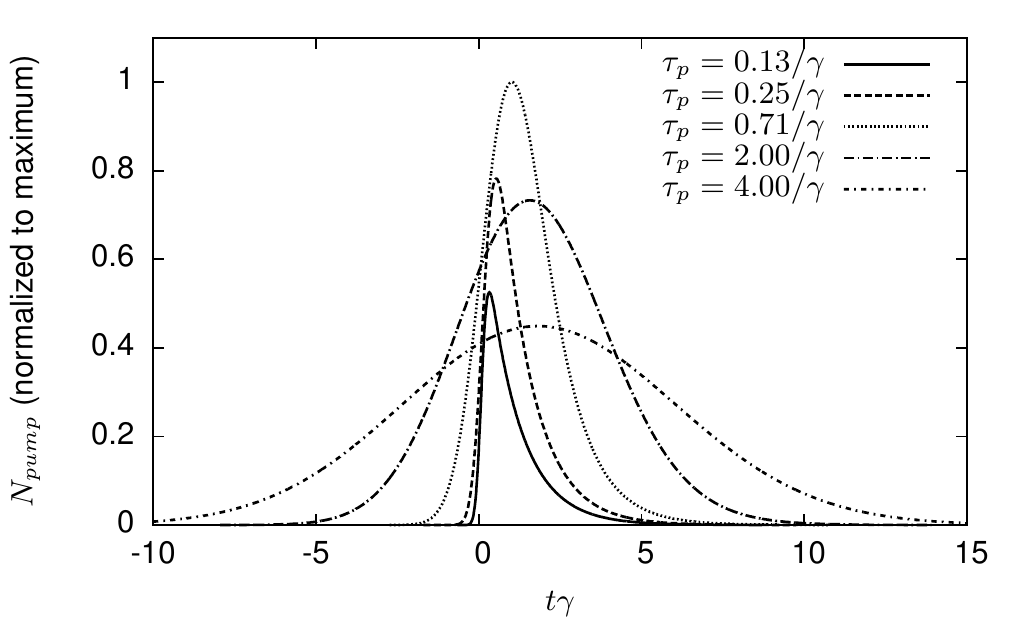}

\includegraphics[width=0.7\columnwidth]{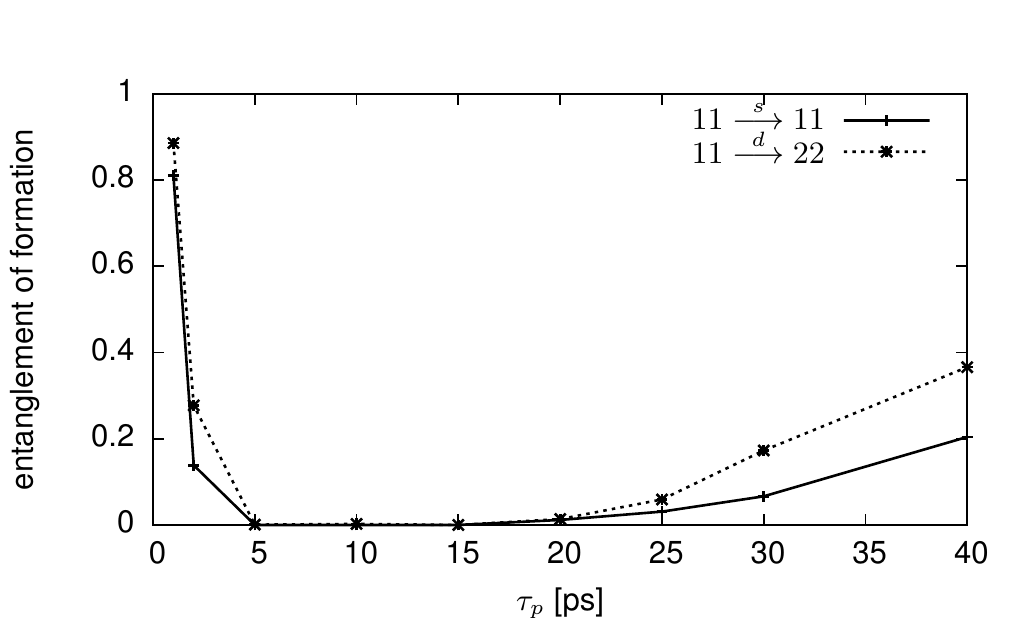}

\includegraphics[width=0.7\columnwidth]{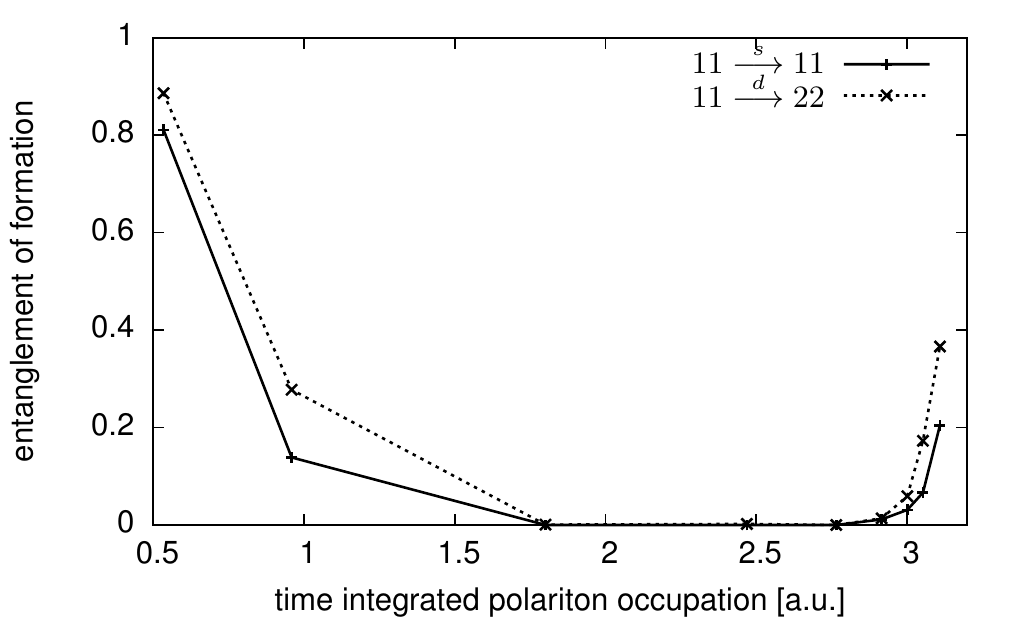}

\caption{The dependence of the polariton population (top) and that of the entanglement
of formation (middle) on the pulse width. The figure on the bottom
is the same as the figure in the middle, with the exception that the
horizontal axis is measured in the total polariton density per pulse
(i.e., integrated over the duration of the excitation pulse.)}
\label{fig:dependence-on-taup} 
\end{figure}

First, we investigate the dependence of the entanglement of formation
on the pump intensity. The results are shown in
figure~\ref{fig:dependence-on-taup}.
Compared to the analytical model, we have set the uniform noise background
to $0$, and only considered pump-induced photoluminescence. Apart
from this fact, the qualitative behaviour of the solutions is similar
to the steady-state case. However, the pump-induced photoluminescence
results in a more rapid decay as a function of the pump intensity.

\begin{figure}
\includegraphics[width=0.95\columnwidth]{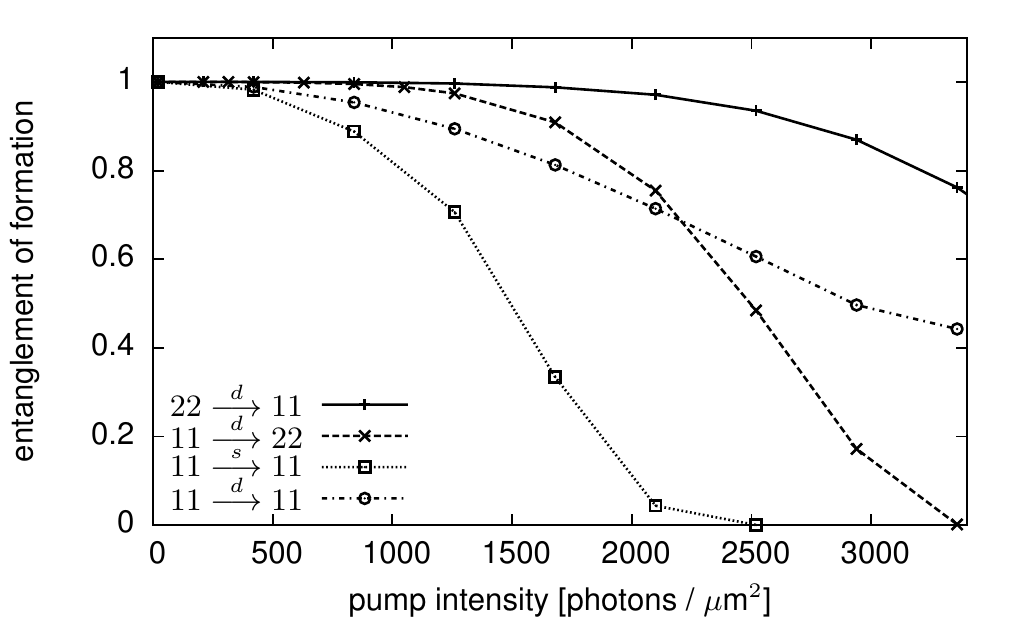}

\label{fig:eof_vs_I}

\caption{Entanglement of formation as a function of the pump intensity (in
units of photons per $\text{\ensuremath{\mu}m}^{2}$) for the \modification{\scheme{11}{11}{s}, \scheme{11}{11}{d},
\scheme{22}{11}{s}, and \scheme{11}{22}{d} pump schemes.}}
\end{figure}

Regarding the various pump schemes, especially in the region of moderate
pump intensities (that are promising for entanglement generation),
the \modification{\scheme{22}{11}{d}}, as well as the \modification{\scheme{11}{22}{d}} schemes are superior to both the
\modification{\scheme{11}{11}{s}}, and the \modification{\scheme{11}{11}{d}} schemes, as demonstrated in
Fig.~\ref{fig:eof_vs_I}. 
This behaviour is likely due to the fact that since signal and idler are located at the same wavevector the effective
population for stimulated polariton scattering is increased. The difference
between the \modification{\scheme{11}{11}{s}} and \modification{\scheme{11}{11}{d}} scheme is the additional factor of
$\frac{1}{2}$ in the coupling coefficient (see Table~\ref{tab:g_k_summary}),
which accounts for most of the discrepancy in the entanglement of
formation. In the double cavity case, the modified Hopfield coefficients
contribute negligibly to $g_{\mathbf{k}}$.

Next, we consider the second parameter of a Gaussian pulse, the pulse
width $\tau_{p}$, which is also of paramount importance, if a highly
entangled state is to be achieved. This is due to the fact that there
exists a \modification{unique peak} close to the decay width of the cavity,
more precisely at $\tau_{p}\approx0.7/\gamma$ for cavity decay width
$\gamma$, such that the peak polariton population is maximized (see
figure~\ref{fig:dependence-on-taup}); this, in turn, has a detrimental
effect on the achievable entanglement of formation. The numerical
simulation shown in figure~\ref{fig:dependence-on-taup} exhibits a
drop in the entanglement of formation in a region around $\tau_{p}\approx\text{\text{10 }ps}$,
as expected from this considerations. For large $\tau_{p}$ our simulation
approximates the steady state case discussed in section \ref{sec:analytical-modeling}.

Therefore, it is advisable to either pump with as narrow pulses as
possible (in schemes were photoluminescence is a concern), or to go
to the steady state case, which is only possible, if a branch protected
from pump-induced photoluminescence is employed.

\begin{figure}
\includegraphics[width=0.95\columnwidth]{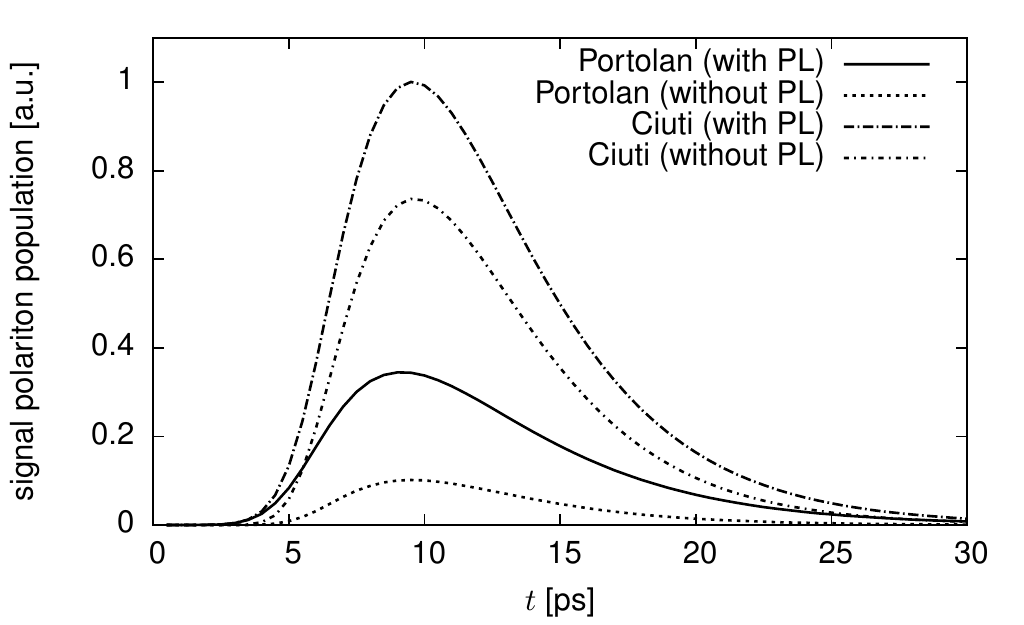}

\caption{Time dependence of the signal polariton population for a pump intensity
of 400 photons/$\mu\mathrm{{m}^{2}}$, pulse width of $1$ ps, and
$\hbar\Omega=3\text{ meV}$. The model of Ciuti et al. gives higher
populations than the model of Portolan et al. both with, and without
background photoluminescence.}

\label{fig:ciutivsportolan}
\end{figure}

The time dependence of the signal polariton population
can be used to highlight the difference between the models of Ciuti
et al., and Portolan et al. This is shown in figure~\ref{fig:ciutivsportolan},
where we fixed the pump wavevectors at $\boldsymbol{k}=(0,0)$ and
$\boldsymbol{k}=(0.9,0.9)\,\mu\text{m}^{-1}$. In addition, we set
the pump intensity to 400 photons/$\mu\mathrm{{m}^{2}}$. The width
of the Gaussian pulse is chosen to be $1$ ps. The pump configuration
corresponds to the \modification{\scheme{11}{11}{s}} scheme. In both cases the signal population
is higher when the photoluminescence is switched on. It is also clear
that either with or without photoluminescence, the population is lower
in the model of Portolan et al. This difference between the two models
is nothing, but the consequence of the difference of the coupling
coefficients, as shown in Table \ref{tab:g_k_summary}. We should
note, however, that the shape of the time evolution in the two models
is approximately the same, and therefore, an experimental verification
of either of them would require the measurement of either absolute
intensities or the investigation of the $k$ dependence.

\section{Conclusions\label{sec:conclusion}}

In this paper, we presented a unified treatment of polariton-polariton scattering in single, double, and triple planar
microcavities, and discussed the selection rules that govern interbranch scattering. Using both a simplified
analytical, and full numerical simulations of the equations of motion, we also investigated how entanglement
in various scattering configurations emerges from a noisy environment caused by background processes. \modification{To
model the pump-induced photoluminescence the simulations have been conducted in the framework of a quantum Langevin
approach. These simulations demonstrated that the double cavity configuration possesses an advantage over single
cavities, because the phonon-induced photoluminescence is somewhat suppressed. In practice, however, additional noise
sources may be present, such as resonant Rayleigh scattering. The double and triple cavity configurations allow us to
choose phase-matching schemes that improve the practical separation of pump and detection beams in an experiment. In
addition, we found that the entanglement of formation} depends quite sensitively on the temporal width of the pump
laser. This is important, because temporal selection of the detected photons is limited by the time resolution of the
photodetectors. The fastest single-photon sensitive photodetectors have a time resolution of several ten picoseconds.
Thus, the temporal optimization can only be done on the excitation side. Finally, we performed a quantitative comparison
of two different approaches to the calculation of the scattering coefficients and we hope that with new experimental
results we will soon be able to decide which approach can better model the dynamics in semiconductor microcavities.

\section*{Acknowledgements}

This work was funded in part by the European Research Council Project No. 257531, and the Canadian Institute for
Advanced Research (CIFAR) through its QIP program. S.P. acknowledges support by the
Austrian Science Fund (FWF) through the START grant Y 591-N16.


\section*{Appendix}

In this part of the paper, we discuss the symmetry properties of the
various polariton states in single, double, and triple planar microcavities
as well as give the interaction terms for the double cavity case.

\subsection{Single cavity}

In the basis of the exciton, $\kex{}$, and the cavity photon, $\kph{}$,
the Hamiltonian of the system is given by 
\begin{equation}
H_{1}=\left(\begin{array}{cc}
E_{x} & \hbar\Omega\\
\hbar\Omega & E_{c}
\end{array}\right)\label{eq:single-cavity-hamiltonian}
\end{equation}
and the eigenstates are the lower and upper polaritons, 
\begin{eqnarray*}
\kp 1 & = & c_{1}\kex{}+c_{2}\kph{}\\
\kp 2 & = & -c_{2}\kex{}+c_{1}\kph{}
\end{eqnarray*}
or in matrix form,
\begin{eqnarray}
\left(\begin{array}{c}
\kp 1\\
\kp 2
\end{array}\right)=\left(\begin{array}{cc}
c_{1} & c_{2}\\
-c_{2} & c_{1}
\end{array}\right)\left(\begin{array}{c}
\kex{}\\
\kph{}
\end{array}\right)\label{eq:transformation-1}
\end{eqnarray}
The eigenvalues take on the form 
\[
E_{j}=\frac{1}{2}\left(E_{c}(\textbf{k})-E_{x}(\textbf{k})+(-1)^{j}\sqrt{(E_{c}(\textbf{k})-
E_{x}(\textbf{k}))^{2}+(2\hbar\Omega)^{2}}\right)\ ,
\]
while the Hopfield coefficients, $c_{1},c_{2}$, are given by the
relation 
\[
\left|c_{j}\right|^{2}=\frac{1}{2}\left(1+(-1)^{j}\frac{E_{c}(\textbf{k})-
E_{x}(\textbf{k})}{\sqrt{(E_{c}(\textbf{k})-E_{x}(\textbf{k}))^{2}+(2\hbar\Omega)^{2}}}\right)\ .
\]
\modification{Note that $c_1$, and $c_2$ represent the excitonic and photonic content of the first polariton branch,
respectively, while the reverse is true for the second branch.}

\subsection{Double cavity}

In the basis of the two excitons, and two photons located in their
respective cavities, the Hamiltonian takes on the form 
\begin{equation}
H_{2}=\left(\begin{array}{cccc}
E_{x_{1}} & \hbar\Omega_{1} & 0 & 0\\
\hbar\Omega_{1} & E_{c_{1}} & 0 & -\hbar\omega_{12}\\
0 & 0 & E_{x_{2}} & \hbar\Omega_{2}\\
0 & -\hbar\omega_{12} & \hbar\Omega_{2} & E_{c_{2}}
\end{array}\right)\label{eq:double-cavity-hamiltonian}
\end{equation}
(Note that the ordering of the basis states is $\kex 1-\kph 1-\kex 2-\kph 2$.)
This Hamiltonian can be made block-diagonal by transformation of the
unitary matrix 
\begin{equation}
U_{2}=U_{2}^{-1}=\frac{1}{\sqrt{2}}\left(\begin{array}{cccc}
1 & 0 & 1 & 0\\
0 & 1 & 0 & 1\\
1 & 0 & -1 & 0\\
0 & 1 & 0 & -1
\end{array}\right)\label{eq:transformation-2}
\end{equation}
i.e., if we use the symmetric-antisymmetric combinations of the two
exciton, and photon states. In the new basis, the Hamiltonian reads
as 
\begin{equation}
\tilde{H}_{2}=U_{2}H_{2}U_{2}^{-1}=\left(\begin{array}{cccc}
E_{x_{1}}(\textbf{k}) & \hbar\Omega_{1} & 0 & 0\\
\hbar\Omega_{1} & E_{c_{1}}(\textbf{k})-\hbar\omega_{12} & 0 & 0\\
0 & 0 & E_{x_{2}}(\textbf{k}) & \hbar\Omega_{2}\\
0 & 0 & \hbar\Omega_{2} & E_{c_{2}}(\textbf{k})+\hbar\omega_{12}
\end{array}\right)\label{eq:h2-block-diagonal}
\end{equation}
and one can readily read off the new eigenstates, which are nothing,
but the eigenstates of the single-cavity Hamiltonian in
equation~(\ref{eq:single-cavity-hamiltonian}),
with a shift in the cavity energies. If we assume, moreover, that
$\Omega=\Omega_{1}=\Omega_{2}$, $E_{x}=E_{x_{1}}=E_{x_{2}}$, $\omega=\omega_{12}$,
and $E_{c}=E_{c_{1}}=E_{c_{2}}$, i.e., that the unperturbed exciton
and cavity energies are degenerate, and that the Rabi splitting does
not depend on the parity of the states, then the eigenvalues of
equation~(\ref{eq:h2-block-diagonal})
take on the particularly simple form 
\[
E_{j}=\frac{E_{c}(\textbf{k})+(-1)^{j}\hbar\omega+E_{x}(\textbf{k})+
(-1)^{\left\lfloor (j+1)/2\right\rfloor }\sqrt{\left(E_{c}(\textbf{k})+(-1)^{j}
\hbar\omega-E_{x}(\textbf{k})\right)^{2}+(2\hbar\Omega)^{2}}}{2}
\]
with the four eigenvectors

\[
\left(\begin{array}{c}
\kp 1\\
\kp 2\\
\kp 3\\
\kp 4
\end{array}\right)=\frac{1}{\sqrt{2}}\left(\begin{array}{cccc}
c_{1} & c_{2} & c_{1} & c_{2}\\
c_{3} & c_{4} & -c_{3} & -c_{4}\\
-c_{2} & c_{1} & -c_{2} & -c_{1}\\
-c_{4} & c_{3} & c_{4} & -c_{3}
\end{array}\right)\left(\begin{array}{c}
\kex 1\\
\kph 1\\
\kex 2\\
\kph 2
\end{array}\right)
\]
where the Hopfield coefficients are given by 
\[
\left|c_{j}\right|^{2}=\frac{1}{2}\left(1+(-1)^{\left\lfloor (j+1)/2\right\rfloor }
\frac{E_{c}(\textbf{k})+(-1)^{j}\hbar\omega-E_{x}(\textbf{k})}{\sqrt{(E_{c}(\textbf{k})+(-1)^{j}
\hbar\omega-E_{x}(\textbf{k}))^{2}+(2\hbar\Omega)^{2}}}\right)
\]

\subsection{Triple cavity }

Ordering the states as $\kex 1-\kph 1-\kex 2-\kph 2-\kex 3-\kph 3$
in the basis of the three excitons, and three photons located in the
three cavities, the Hamiltonian reads as 
\begin{equation}
H_{3}=\left(\begin{array}{cccccc}
E_{x_{1}}(\textbf{k}) & \hbar\Omega_{1} & 0 & 0 & 0 & 0\\
\hbar\Omega_{1} & E_{c_{1}}(\textbf{k}) & 0 & -\hbar\omega_{12} & 0 & 0\\
0 & 0 & E_{x_{2}}(\textbf{k}) & \hbar\Omega_{2} & 0 & 0\\
0 & -\hbar\omega_{12} & \hbar\Omega_{2} & E_{c_{2}}(\textbf{k}) & 0 & -\hbar\omega_{23}\\
0 & 0 & 0 & 0 & E_{x_{3}} & \hbar\Omega_{3}\\
0 & 0 & 0 & -\hbar\omega_{23} & \hbar\Omega_{3} & E_{c_{3}}
\end{array}\right)\label{eq:triple-cavity-hamiltonian}
\end{equation}

Again, with the simplifying assumption $\Omega=\Omega_{1}=\Omega_{2}=\Omega_{3}$,
$E_{x}=E_{x_{1}}=E_{x_{2}}=E_{x_{3}}$, $\omega=\omega_{12}=\omega_{23}$,
and $E_{c}=E_{c_{1}}=E_{c_{2}}=E_{c_{3}}$, the transformation matrix
\begin{equation}
U_{3}=\frac{1}{\sqrt{2}}\left(\begin{array}{cccccc}
1 & 0 & 0 & 0 & -1 & 0\\
0 & 1 & 0 & 0 & 0 & -1\\
\frac{1}{\sqrt{2}} & 0 & 1 & 0 & \frac{1}{\sqrt{2}} & 0\\
0 & \frac{1}{\sqrt{2}} & 0 & 1 & 0 & \frac{1}{\sqrt{2}}\\
\frac{1}{\sqrt{2}} & 0 & -1 & 0 & \frac{1}{\sqrt{2}} & 0\\
0 & \frac{1}{\sqrt{2}} & 0 & -1 & 0 & \frac{1}{\sqrt{2}}
\end{array}\right)\label{eq:transformation-3}
\end{equation}
brings $H_{3}$ into a block-diagonal form, and we get 
\begin{equation}
\tilde{H}_{3}=U_{3}H_{3}U_{3}^{-1}=\left(\begin{array}{cccccc}
E_{x} & \hbar\Omega & 0 & 0 & 0 & 0\\
\hbar\Omega & E_{c}-\sqrt{2}\hbar\omega & 0 & 0 & 0 & 0\\
0 & 0 & E_{x} & \hbar\Omega & 0 & 0\\
0 & 0 & \hbar\Omega & E_{c} & 0 & 0\\
0 & 0 & 0 & 0 & E_{x} & \hbar\Omega\\
0 & 0 & 0 & 0 & \hbar\Omega & E_{c}+\sqrt{2}\hbar\omega
\end{array}\right)\label{eq:h3-block-diagonal}
\end{equation}
with the eigenvalues 
\begin{align*}
E_{j}=&\frac{E_{c}(\textbf{k})-2\hbar\omega\cos\left(j\frac{\pi}{4}\right)+
E_{x}(\textbf{k})}{2}
\\
&\quad + \frac{(-1)^{\left\lfloor \frac{j+2}{3}\right\rfloor
}\sqrt{\left(E_{c}(\textbf{k})-2\hbar\omega\cos\left(j\frac{\pi}{4}\right)-E_{x}
(\textbf{k} )\right){}^{2}+(2\hbar\Omega)^{2}}}{2}
\end{align*}
The polariton eigenstates are

\[
\left(\begin{array}{c}
\kp 1\\
\kp 2\\
\kp 3\\
\kp 4\\
\kp 5\\
\kp 6
\end{array}\right)=\left(\begin{array}{cccccc}
\frac{c_{1}}{\sqrt{2}} & \frac{c_{2}}{\sqrt{2}} & 0 & 0 & -\frac{c_{1}}{\sqrt{2}} & -\frac{c_{2}}{\sqrt{2}}\\
-\frac{c_{2}}{\sqrt{2}} & \frac{c_{1}}{\sqrt{2}} & 0 & 0 & \frac{c_{2}}{\sqrt{2}} & -\frac{c_{1}}{\sqrt{2}}\\
\frac{c_{3}}{2} & \frac{c_{4}}{2} & \frac{c_{3}}{\sqrt{2}} & \frac{c_{4}}{\sqrt{2}} & \frac{c_{3}}{2} & \frac{c_{4}}{2}\\
-\frac{c_{4}}{2} & \frac{c_{3}}{2} & -\frac{c_{4}}{\sqrt{2}} & \frac{c_{3}}{\sqrt{2}} & -\frac{c_{4}}{2} & \frac{c_{3}}{2}\\
\frac{c_{5}}{2} & \frac{c_{6}}{2} & -\frac{c_{5}}{\sqrt{2}} & -\frac{c_{6}}{\sqrt{2}} & \frac{c_{5}}{2} & \frac{c_{6}}{2}\\
-\frac{c_{6}}{2} & \frac{c_{5}}{2} & \frac{c_{6}}{\sqrt{2}} & -\frac{c_{5}}{\sqrt{2}} & -\frac{c_{6}}{2} & \frac{c_{5}}{2}
\end{array}\right)\left(\begin{array}{c}
\kex 1\\
\kph 1\\
\kex 2\\
\kph 2\\
\kex 3\\
\kph 3
\end{array}\right)
\]
where we have simplified the matrix by noting that $c_{1}^{2}+c_{2}^{2}=c_{3}^{2}+c_{4}^{2}=c_{5}^{2}+c_{6}^{2}=1$.
$c_{j}$ are the Hopfield coefficients of the three block-diagonals in equation~(\ref{eq:transformation-3}), and are
given by

\[
\left|c_{j}\right|^{2}=\frac{1}{2}\left(1+(-1)^{\left\lfloor \frac{j+2}{3}\right\rfloor }
\frac{E_{c}(\textbf{k})-2\hbar\omega\cos\left(j\frac{\pi}{4}\right)-E_{x}(\textbf{k})}
{\sqrt{\left(E_{c}(\textbf{k})-2\hbar\omega\cos\left(j\frac{\pi}{4}\right)-E_{x}(\textbf{k})\right){}^{2}
+(2\hbar\Omega)^ {2}}}\right).
\]

By applying the transformation, we changed the description of the
problem from the basis of excitons and photons located in their respective
cavity to the basis of the totally antisymmetric, $\kex{as}=\frac{1}{\sqrt{2}}\mbox{\ensuremath{\left(\kex 1-\kex 3\right)}}$,
and totally symmetric $\kex{s_{1}}=\frac{1}{2}\mbox{\ensuremath{\kex 1+\frac{1}{\sqrt{2}}\kex 2}+\ensuremath{\frac{1}{2}\kex 3}}$,
$\kex{s_{2}}=\frac{1}{2}\mbox{\ensuremath{\kex 1-\frac{1}{\sqrt{2}}\kex 2}+\ensuremath{\frac{1}{2}\kex 3}}$
wavefunctions. (Similar expressions apply to the photon states.)

\subsection{The interaction terms in the polariton basis\label{sub:interaction-terms}}

\begin{sideways}
\parbox{\textheight}{
\begin{align}
R_{\vk}^{\mathrm{xx,1}} & =V_{\mathrm{xx}}\sum_{\vk_{1},\vk_{2}}B_{1\vk_{1}+\vk_{2}-\vk}^{\dagger}B_{1\bld k_{1}}B_{1\bld k_{2}}\nonumber \\
 & =\frac{V_{\mathrm{xx}}}{2\sqrt{2}}\sum_{\vk_{1},\vk_{2}}\biggl\{\left(c_{1\vk_{1}+\vk_{2}-\vk}p_{1\vk_{1}+\vk_{2}-\vk}^{\dagger}+
c_{3\vk_{1}+\vk_{2}-\vk}p_{2\vk_{1}+\vk_{2}-\vk}^{\dagger}-c_{2\vk_{1}+\vk_{2}-\vk}p_{3\vk_{1}+\vk_{2}-\vk}^{\dagger}-
c_{4\vk_{1}+\vk_{2}-\vk}p_{4\vk_{1}+\vk_{2}-\vk}^{\dagger}\right)\times\nonumber \\
 & \qquad\left(c_{1\bld k_{1}}p_{1\bld k_{1}}+c_{3\bld k_{1}}p_{2\bld k_{1}}-c_{2\bld k_{1}}p_{3\bld k_{1}}-
c_{4\bld k_{1}}p_{4\bld k_{1}}\right)\left(c_{1\bld k_{2}}p_{1\bld k_{2}}+c_{3\bld k_{2}}p_{2\bld k_{2}}-
c_{2\bld k_{2}}p_{3\bld k_{2}}-c_{4\bld k_{2}}p_{4\bld k_{2}}\right)\biggr\}\label{eq:R_xx_1-1}
\end{align}
\begin{align}
R_{\vk}^{\mathrm{xx,2}} & =V_{\mathrm{xx}}\sum_{\vk_{1},\vk_{2}}B_{2\vk_{1}+\vk_{2}-\vk}^{\dagger}B_{2\bld
k_{1}}B_{2\bld k_{2}}\nonumber \\
 & =\frac{V_{\mathrm{xx}}}{2\sqrt{2}}\sum_{\vk_{1},\vk_{2}}\biggl\{\left(c_{1\vk_{1}+\vk_{2}-\vk}p_{1\vk_{1}+\vk_{2}-\vk}^{\dagger}-
c_{3\vk_{1}+\vk_{2}-\vk}p_{2\vk_{1}+\vk_{2}-\vk}^{\dagger}-c_{2\vk_{1}+\vk_{2}-\vk}p_{3\vk_{1}+\vk_{2}-\vk}^{\dagger}+
c_{4\vk_{1}+\vk_{2}-\vk}p_{4\vk_{1}+\vk_{2}-\vk}^{\dagger}\right)\times\nonumber \\
 & \qquad\left(c_{1\bld k_{1}}p_{1\bld k_{1}}-c_{3\bld k_{1}}p_{2\bld k_{1}}-c_{2\bld k_{1}}p_{3\bld k_{1}}+
c_{4\bld k_{1}}p_{4\bld k_{1}}\right)\left(c_{1\bld k_{2}}p_{1\bld k_{2}}-c_{3\bld k_{2}}p_{2\bld k_{2}}-
c_{2\bld k_{2}}p_{3\bld k_{2}}+c_{4\bld k_{2}}p_{4\bld k_{2}}\right)\biggr\}\label{eq:R_xx_2-1}
\end{align}
}
\end{sideways}
\clearpage
\begin{sideways}
\parbox{\textheight}{
\begin{align}
R_{\bld k}^{\mathrm{sat},1} & =\frac{V}{n_{\mathrm{sat}}}\sum_{\bld k_{1},\bld k_{2}}
B_{1\bld k_{1}}a_{1\bld k_{2}}B_{1\vk_{1}+\vk_{2}-\vk}^{\dagger}\nonumber \\
 & =\frac{V}{2\sqrt{2}n_{\mathrm{sat}}}\sum_{\vk_{1},\vk_{2}}\biggl\{\left(c_{1\bld k_{1}}
p_{1\bld k_{1}}+c_{3\bld k_{1}}p_{2\bld k_{1}}-c_{2\bld k_{1}}p_{3\bld k_{1}}-c_{4\bld k_{1}}
p_{4\bld k_{1}}\right)\left(c_{2\bld k_{2}}p_{1\bld k_{2}}+c_{4\bld k_{2}}p_{2\bld k_{2}}+
c_{1\bld k_{2}}p_{3\bld k_{2}}+c_{4\bld k_{2}}p_{2\bld k_{2}}\right)\times\nonumber \\
 & \qquad\left(c_{1\vk_{1}+\vk_{2}-\vk}p_{1\vk_{1}+\vk_{2}-\vk}^{\dagger}+
c_{3\vk_{1}+\vk_{2}-\vk}p_{2\vk_{1}+\vk_{2}-\vk}^{\dagger}-c_{2\vk_{1}+\vk_{2}-\vk}
p_{3\vk_{1}+\vk_{2}-\vk}^{\dagger}-c_{4\vk_{1}+\vk_{2}-\vk}p_{4\vk_{1}+\vk_{2}-\vk}^{\dagger}\right)\biggr\}
\label{eq:R_sat_1-1}
\end{align}
\begin{align}
R_{\vk}^{\mathrm{sat},2} & =\frac{V}{n_{\mathrm{sat}}}\sum_{\vk_{1},\vk_{2}}B_{2\bld k_{1}}
a_{2\bld k_{2}}B_{2\vk_{1}+\vk_{2}-\vk}^{\dagger}\nonumber \\
 & =\frac{V}{2\sqrt{2}n_{\mathrm{sat}}}\sum_{\vk_{1},\vk_{2}}\biggl\{\left(c_{1\bld k_{1}}
p_{1\bld k_{1}}-c_{3\bld k_{1}}p_{2\bld k_{1}}-c_{2\bld k_{1}}p_{3\bld k_{1}}+c_{4\bld k_{1}}
p_{4\bld k_{1}}\right)\left(c_{2\bld k_{2}}p_{1\bld k_{2}}-c_{4\bld k_{2}}p_{2\bld k_{2}}+
c_{1\bld k_{2}}p_{3\bld k_{2}}-c_{4\bld k_{2}}p_{2\bld k_{2}}\right)\times\nonumber \\
 & \qquad\left(c_{1\vk_{1}+\vk_{2}-\vk}p_{1\vk_{1}+\vk_{2}-\vk}^{\dagger}-c_{3\vk_{1}+
\vk_{2}-\vk}p_{2\vk_{1}+\vk_{2}-\vk}^{\dagger}-c_{2\vk_{1}+\vk_{2}-\vk}p_{3\vk_{1}+
\vk_{2}-\vk}^{\dagger}+c_{4\vk_{1}+\vk_{2}-\vk}p_{4\vk_{1}+\vk_{2}-\vk}^{\dagger}\right)\biggr\}.
\label{eq:R_sat_2-1}
\end{align}
}
\end{sideways}

\subsection{A simple formula for the entanglement of formation \label{sub:approx-eof}}

\modification{The entanglement of formation, $E(\rho)$, is one of the most commonly employed measures of entanglement
found in the literature; it quantifies the resources needed to create a given entangled state. This is discussed at
length in \cite{Hill1997,Wootters1998}.} For a density matrix $\rho$, $E(\rho)$ is given by

\[
E(\rho)=h\left(\frac{1+\sqrt{1-C(\rho)^{2}}}{2}\right)
\]
where $h$ is the entropy 
\[
h(x)=-x\log_{2}x-(1-x)\log_{2}(1-x)
\]
and 
\[
C(\rho)=\max\left(0,\lambda_{1}-\sum_{i\geq2}\lambda_{i}\right),
\]
where $\lambda_{1}>\lambda_{2}>\dots>\lambda_{n}$ are the ordered
eigenvalues of the operator 
\[
\sqrt{\sqrt{\rho}\left(\sigma_{y}\otimes\sigma_{y}\right)\rho^{*}\left(\sigma_{y}\otimes\sigma_{y}\right)\sqrt{\rho}}\ .
\]
\modification{Applying the formulae above and equation~(\ref{eq:density-matrix-steady-state}), it is straightforward to
show that in the steady state}

\[
C(\rho)=\begin{cases}
0 & \Delta\leq n\\
\frac{\left(4\Delta^{2}-1\right)(n-\Delta)(\Delta+n)}{8\Delta^{4}+2n^{2}+\Delta^{2}(4n(n+3)+1)} & \text{otherwise}
\end{cases}\ .
\]
In order to get a simplified expression for the EOF, we choose to
approximate the expression for the binary entropy by the linear function
$h(x)\approx0.065+0.98x$ (which is the best linear fit in the infinity
norm), resulting in 
\[
E(\rho)\approx0.065+\frac{0.98\left(n^{2}-\Delta^{2}\right)^{2}}{\left(2n^{2}+(1+12n)\Delta^{2}+8\Delta^{4}\right)^{2}}.
\]
This captures the essential dynamics of the EOF as shown in
figure~\ref{fig:eof_steady_state}.

\bibliographystyle{unsrt}
\bibliography{polariton}

\end{document}